\documentclass[iop]{emulateapj}
\usepackage{natbib}
\usepackage{epsfig}
\usepackage{subfigure}
\usepackage{graphics}

\newcommand{\chan}{{\it Chandra}}

\newcommand{\kms}{$\,\rm{km\,s^{-1}}$}
\newcommand{\msolar}{$\rm{M}_{\odot}$}
\shorttitle{GAMA and the assembly of A1882}
\shortauthors{M.S. Owers {\it et al.}}
\begin{document}
\title{Galaxy and Mass Assembly (GAMA): Witnessing the assembly of the cluster Abell~1882}


\author{M. S. Owers\altaffilmark{1,12}, 
I. K. Baldry\altaffilmark{2}, 
A. E. Bauer\altaffilmark{1},
J. Bland-Hawthorn\altaffilmark{3},
M. J. I. Brown\altaffilmark{4},
M. E. Cluver\altaffilmark{1},
M. Colless\altaffilmark{5}, 
S. P. Driver\altaffilmark{6,7},
A. C. Edge\altaffilmark{8}, 
A. M. Hopkins\altaffilmark{1}, 
E. van Kampen\altaffilmark{9},
M. A. Lara-Lopez\altaffilmark{1},
J. Liske\altaffilmark{10},
J. Loveday\altaffilmark{10},
K. A. Pimbblet\altaffilmark{3}, 
T. Ponman\altaffilmark{11},
A. S. G. Robotham\altaffilmark{5,6}}
\altaffiltext{1}{Australian Astronomical Observatory, PO Box 915, North Ryde, NSW 1670, Australia}
\altaffiltext{2}{Astrophysics Research Institute, Liverpool John Moores University, UK} 
\altaffiltext{3}{Sydney Institute for Astronomy, School of Physics A28, University of Sydney, NSW 2006, Australia} 
\altaffiltext{4}{School of Physics, Monash University, Clayton, Vic 3800, Australia} 
\altaffiltext{5}{Research School of Astronomy \& Astrophysics, Australian National University, Australia} 
\altaffiltext{6}{School of Physics \& Astronomy, University of St Andrews, North Haugh, St Andrews KY16 9SS, UK} 
\altaffiltext{7}{International Center for Radio Astronomy Research (ICRAR), 
The University of Western Australia, 35 Stirling Highway, Crawley, WA 6009, Australia} 
\altaffiltext{8}{Institute of Computational Cosmology, Durham University, Durham DH1 3LE, UK} 
\altaffiltext{9}{European Southern Observatory, Karl-Schwarzschild-Str. 2, 85748 Garching, Germany} 
\altaffiltext{10}{Astronomy Center, University of Sussex, Falmer, Brighton BN1 9QH, UK} 
\altaffiltext{11}{School of Physics \& Astronomy, University of Birmingham, Birmingham B15 2TT, UK} 
\altaffiltext{12}{Australian Research Council Super Science Fellow; E-mail:mowers@aao.gov.au}

\begin{abstract}
We present a combined optical and X-ray analysis of the rich cluster Abell~1882 with the aim of identifying merging substructure and understanding the recent assembly history of this system. Our optical data consist of spectra drawn from the Galaxy and Mass Assembly (GAMA) survey, which lends itself to this kind of detailed study thanks to its depth and high spectroscopic completeness. We use 283 spectroscopically confirmed cluster members to detect and characterize substructure. We complement the optical data with X-ray data taken with both \chan\ and XMM. Our analysis reveals that A1882 harbors two main components, A1882A and A1882B, which have a projected separation of $\sim 2\,$Mpc and a line of sight velocity difference of $v_{los} \sim -428^{+187}_{-139}\,$\kms. The primary system, A1882A, has velocity dispersion $\sigma_v=500_{-26}^{+23}\,$\kms\, and \chan\ (XMM) temperature $kT=3.57\pm0.17\,$keV ($3.31^{+0.28}_{-0.27}\,$keV) while the secondary, A1882B, has $\sigma_v=457^{+108}_{-101}\,$\kms\, and \chan\ (XMM) temperature $kT=2.39\pm0.28\,$keV ($2.12\pm0.20\,$keV). The optical and X-ray estimates for the masses of the two systems are consistent within the uncertainties and indicate that there is twice as much mass in A1882A ($M_{500}=1.5-1.9 \times 10^{14}\,$\msolar) when compared with A1882B ($M_{500}=0.8-1.0 \times 10^{14}\,$\msolar). We interpret the A1882A/A1882B system as being observed prior to a core passage. Supporting this interpretation is the large projected separation of A1882A and A1882B and the dearth of evidence for a recent ($<2\,$Gyr) major interaction in the X-ray data. Two-body analyses indicate that A1882A and A1882B form a bound system with bound incoming solutions strongly favored. We compute blue fractions of $f_b=0.28\pm0.09$ and $0.18\pm0.07$ for the spectroscopically confirmed member galaxies within $r_{500}$ of the centers of A1882A and A1882B, respectively. These blue fractions do not differ significantly from the blue fraction measured from an ensemble of 20 clusters with similar mass and redshift.
\end{abstract}


\section{Introduction}

The observed properties of the large scale structure in our Universe are well described by a $\Lambda$CDM cosmological model \citep{springel2006}. Within this model the formation of structure progresses in a hierarchical fashion, culminating with the formation of clusters of galaxies. Hierarchical cluster growth occurs via several modes with varying degrees of impact on the state of the cluster, from the benign continuous infall of material from the surrounding filaments, to the high impact merger of two approximately equal mass clusters. Simulations indicate that a significant fraction of both the mass and galaxies in massive ($\sim 10^{14-15}\,$\msolar) clusters at the current epoch have been accreted through minor and major cluster mergers \citep[$\sim 40-50\%$,][]{berrier2009,mcgee2009}. Therefore, it is important that we understand the impact of this process on the cluster constituents and, in particular, how this violent environment affects the resident galaxies.

Initial indications that cluster mergers may affect the star formation in the resident galaxies came from observations of the Coma cluster, where \citet{caldwell1993} discovered an excess of rapidly-evolving post-starburst galaxies coincident with a merging subgroup to the southwest of the cluster core. Further investigation by \citet{poggianti2004} revealed that galaxies with evidence for recently truncated episodes of starburst activity were co-spatial with intra-cluster medium (ICM) substructures associated with the dynamical evolution of the cluster. This indicates that an interaction with the dynamically evolving ICM may be responsible for the triggering and/or truncation of star formation in these galaxies. Simulations support this conclusion and show that it is possible that the high ICM pressure a galaxy experiences during the core-passage phase of a merger can trigger star formation \citep{roettiger1996, bekki2003, kronberger2008, bekki2010} while the high relative velocity of ICM and galaxies can enhance ram pressure stripping of the interstellar medium, leading to a sharp truncation of star formation \citep{fujita1999}. Observations at optical and radio wavelengths of several other merging clusters support this scenario \citep[][]{caldwell1997, venturi2000, venturi2001, venturi2002, miller2003, giacintucci2004, miller2006, johnstonhollitt2008, hwang2009, ma2010, owers2012}. Since the timescales for the radio and star forming phases of galaxies ($1-100\,$Myr) are shorter than typical merger timescales ($\sim\,$Gyrs), a detailed understanding of the dynamics and merger stage of the cluster are crucial when attempting to interpret the observed galaxy populations.

The combination of multi-object spectroscopy with X-ray spectro-imagery has proven a powerful tool in understanding cluster mergers \citep{owers2009a, owers2011a, maurogordato2008, maurogordato2011, ma2009, ma2010, barrena2007}. The multi-object spectroscopy allows efficient collection of large, highly complete, samples of spectroscopically confirmed cluster member galaxies. These member galaxies act as excellent kinematic probes that can be used to first identify merger related substructures, and then to determine substructure characteristics. These are important for constraining merger configurations, such as velocity dispersion and the line of sight velocity with respect to the parent cluster. High fidelity X-ray data, such as that provided by the \chan\, and XMM-Newton satellites, maps the distribution and thermodynamic properties of the ICM. The collisional nature of the ICM means that it provides a number of morphological and thermodynamic signatures of merger activity such as shocks \citep{markevitch2002, markevitch2005, russell2010, marcario2011, owers2011a} and cold fronts \citep{markevitch2000, vikhlinin2001b, markevitch2007, owers2009c}. These signatures are extremely useful in inferring the direction of motion of structures \citep{maurogordato2011}, the merger velocity perpendicular to our line of sight \citep{markevitch2002}, and also for understanding if a merger is observed at pre- or post-pericentric passage. The complementary nature of these two probes of cluster mergers allows tight constraints to be placed on merger configurations and histories, allowing a more complete understanding of the merger process and the identification of regions which are currently, or have recently been, affected by the cluster merger.

In this paper we present a detailed analysis of the cluster Abell~1882 (hereafter A1882) utilizing the highly complete GAMA spectroscopic data along with archival \chan\ and XMM data. A1882 is the richest cluster in the GAMA group catalog \citep{robotham2011} where it was allocated 264 members, median redshift z=0.1394 and velocity dispersion $\sigma=833\,$\kms. It is an Abell richness class 3 \citep[][]{abell1958} and was included in the \citet{morrison2003} multiwavelength study of rich Abell clusters where an X-ray luminosity $\rm L_{X}(0.5-2.0 {\rm keV}) = 3.02\times10^{43}\,$ erg\,${\rm s}^{-1}$ was measured. A1882 was notable in this study as having the highest fraction of blue galaxies, $f_b=0.28$. The isopleth maps presented there showed a complex multi-modal distribution in galaxy surface density, while the X-ray images revealed multiple peaks in the ICM distribution, indicating that A1882 is not a relaxed system and may be undergoing a merger. However, these substructures may simply be due to fore- and background structures aligned along the line of sight that are not physically associated with A1882. Moreover, the low resolution X-ray images used in \citet{morrison2003} are prone to point-source contamination while the projection-effect-prone isopleth maps give little information on the details of the merger. These merger details are necessary to understand the impact of cluster mergers on the member galaxies and cannot be achieved by shallow, large-area surveys which do not obtain high spectroscopic completeness in dense environments. 

The aim of this paper is to answer two questions: (i) What is the dynamical state of A1882; is it in a pre- or post core passage merger phase? and (ii) What is the nature of the apparently high blue fraction within A1882 and is it anomalous? The first question is addressed by using a sample of spectroscopically confirmed cluster members selected from the GAMA survey, along with archival \chan\ and XMM data, to detect and characterize substructure. The high spectroscopic completeness ($\sim 99\%$ even in dense cluster environments) and depth ($r < 19.8$) of the GAMA survey is crucial to allow the robust identification and characterization of dynamical substructure, which is usually only attainable through pointed observations. To address the second question, we make use of the GAMA Group Catalog \citep{robotham2011} to select a benchmark sample of mass- and redshift-matched clusters for comparing blue fractions. This study forms part of a larger body of work aimed at understanding the impact of hierarchical structure formation on cluster galaxies. In previous studies, we have provided detailed pictures of the merger states of several clusters ranging from post-core passage major mergers \citep{owers2009b, owers2011a, owers2012} to minor mergers first identified by the existence of cold fronts \citep{owers2009a, owers2009c, owers2011b}. In a forthcoming paper, we will compare the galaxy properties across the spectrum of cluster dynamical states in order to asses the effect of the merger induced rapidly changing environment.

In Section~\ref{data} we describe the GAMA, \chan\ and XMM data used in this study. In Section~\ref{optical} we present the analysis of the optical data which includes determination of cluster membership and techniques used for the detection of substructure. In Section~\ref{xray} we present the X-ray analysis. In Section~\ref{discussion} we determine subcluster masses, discuss merger scenarios and determine whether the blue fraction in A1882 is truly anomalous. We summarize our results and present conclusions in Section~\ref{summary}. Throughout this paper, we assume a standard $\Lambda{\rm CDM}$ cosmology with $H_0=70\,$\kms ${\rm Mpc}^{-1}$, $\Omega_m=0.3$ and $\Omega_{\Lambda}=0.7$. For the assumed cosmology and at the cluster redshift ($z=0.1389$; Section~\ref{vpec_dist}) $1''=2.45\,$kpc.

\section{Data}\label{data}
\subsection{GAMA data}

GAMA\footnote{www.gama-survey.org/} is a multi-wavelength data endeavor built around a highly complete ($99\%$) spectroscopic survey of $\sim 240,000$ galaxies to a limiting magnitude of $r = 19.8$ \citep{driver2009, driver2011}. The majority (around $85\%$) of the spectra were taken at the 3.9m Anglo Australian telescope with the AAOmega instrument. AAOmega is a bench-mounted, dual-beam spectrograph fed by 392 fibers which are positioned on the prime-focus-mounted Two Degree Field instrument \citep{saunders2004, smith2004, sharp2006}. The target selection is described in detail in \citet{baldry2010}, the tiling in \citet{robotham2010}, the instrument configuration, exposure times and redshift measurement details in \citet{driver2011} while the data processing is described in \citet{hopkins2013}. The majority of the remaining $\sim 15\%$ of spectra come from the Two-degree Field Galaxy Redshift Survey \citep{colless2001} and the SDSS DR7 \citep{abazajian2009} with the remainder coming from sources listed in \citet[][]{driver2011}. In this paper we utilize only a small portion of the spectroscopic redshifts (drawn from SpecCatv17 in the GAMA-II survey), specifically, those found within a $24$ arcminute radius centered on the brightest cluster galaxy in A1882 (R.A.=$14:15:08.39$, Decl.=$-00:29:35.7$).

\subsection{Archival X-ray data}

We use archival X-ray observations of Abell~1882 taken with XMM-Newton using the European Photon Imaging Camera (EPIC) in February 2003 (ObsID 0145480101) and with the Advanced CCD Imaging Spectrometer (ACIS) onboard \chan\ in March, September and December 2011. The EPIC observations were performed in full-frame mode with the medium filter for a total exposure times of 23.3\,ks and 21.7\,ks for the MOS (Metal Oxide Semi-conductor) and PN CCD arrays, respectively, centered at R.A.=14:14:48.0, decl.=-00:24:00.0. The nine \chan\ pointings used the ACIS-S array and were centered on the back-illuminated S3 chip and were taken in VFAINT data mode. The \chan\ observations are summarized in Table~\ref{xray_obs}.

\begin{deluxetable}{ccccc}
  \tablecaption{Summary of the nine \chan\ X-ray pointings. \label{xray_obs}} 
  \tablecolumns{5}
\tablehead {\colhead{ObsIDs}   & \colhead{R.A.} & \colhead{decl.} & \colhead{$T_{exp}$} &  \colhead{Cleaned $T_{exp}$}\\
& & &  \colhead{(ks)} & \colhead{(ks)}}\\
\startdata
12904 &  14:15:06.60 &	$-00:29:27.60$ & 32.94 & 30.61 \\
12905 &  14:15:06.60 &	$-00:29:27.60$ & 32.94 & 30.89 \\
12906 &  14:15:06.60 &	$-00:29:27.60$ & 32.94 & 29.37 \\
12907 &  14:14:24.50 &	$-00:22:37.90$ & 13.20 & 12.26 \\
12908 &  14:14:24.50 &	$-00:22:37.90$ & 12.93 & 12.93 \\
12909 &  14:14:24.50 &	$-00:22:37.90$ & 13.20 & 12.18 \\
12910 &  14:14:57.90 &	$-00:20:55.70$ & 16.49 & 16.49 \\
12911 &  14:14:57.90 &	$-00:20:55.70$ & 16.23 & 16.23 \\
12912 &  14:14:57.90 &	$-00:20:55.70$ & 16.22 & 15.20 \\
\enddata
\end{deluxetable}

\subsubsection{XMM-Newton}
The XMM Observation Data Files (ODF) are reprocessed using the XMM-Newton Science Analysis Software (SAS; version 12.01) tasks {\sf emchain} and {\sf ephain} for the MOS and PN data, respectively. The data are filtered for periods of high background due to soft proton flares with the {\sf espfilt} task. Roughly $50\%$ of the MOS observations were rejected due to flare contamination leaving a cleaned exposure times of 11.0\,ks and 11.6\,ks for MOS1 and MOS2, respectively. The PN data were severely affected by flares, with roughly 70 percent rejected as being contaminated by flares, leaving 6.9\,ks of clean exposure.

{ For the XMM imaging analysis, we make use of blank sky and filter wheel closed (FWC) observations produced by the EPIC background team and tailored to the observations\footnote{http://xmm.vilspa.esa.es/external/xmm\_sw\_cal/background/index.shtml} \citep{carter2007}. These datasets are filtered to exclude periods of high background evident in the 10--12\,keV and 2--7\,keV band light curves. We use the {\sf imagBGsub} software\footnote{http://www.sr.bham.ac.uk/xmm3/scripts.html} to produce background corrected images using a double background subtraction procedure. Briefly, this method uses the FWC observations to subtract the instrumental background from both the blank sky observations and a source free region in the observations leaving only the cosmic X-ray background. Due to differences in sky pointings between the observations and blank sky datasets, there are small differences in the soft X-ray background flux. This is accounted for by comparing the cosmic X-ray background flux in the blank sky with that in a source-free region in the observations. The comparison is made in four energy bands in the 0.5--2.5\,keV range with the differences used to make vignetting-corrected ``soft excess'' images. These soft-excess images are combined with instrumental and cosmic X-ray background images to produce a total background which is subtracted from the observations. Images are binned to have $4\arcsec \times 4\arcsec$ pixels and are restricted to the 0.5--7\,keV energy range. Corresponding exposure maps which correct for vignetting are also produced. }

Spectral analyses are performed in the 0.5--7\,keV energy range. Auxiliary response files (ARF), which correct filter transmission, quantum efficiency, effective area are generated with the SAS task {\sf arfgen}. Redistribution matrix files (RMF) which describe the response as a function of energy, are generated with the SAS task {\sf rmfgen}. For background subtraction, we use blank sky observations produced by the EPIC background team and tailored to the observations\footnote{xmm.vilspa.esa.es/external/xmm\_sw\_cal/background} \citep{carter2007}. To account for the soft background excess due to differences in sky pointings between the observations and blank sky datasets, we extract spectra and responses for an annular region which is free of source emission. We extract a background from the same region of the blank sky observations.  We use XSPEC to simultaneously fit residual soft X-ray background emission for all three cameras with two unabsorbed, redshift zero, solar metallicity MEKAL models. The best fitting temperatures were found to be kT=$0.17\pm0.02\,$keV and kT=$0.59\pm0.02\,$keV. This background model, corrected for the ratio of the extraction region areas, is included in determining the mean temperatures presented in Section~\ref{temps}. The inclusion of this extra background increases the measured temperature by $\sim 10\%$.

\subsubsection{\chan}\label{chandra_red}
The \chan\ level 1 data were reprocessed using the {\sf chandra\_repro} tool within the {\it CIAO} package  \citep[version 4.4;][]{fruscione2006} with the latest gain and calibration files applied (CalDB version 4.4.7) and VFAINT background cleaning applied. Light curves were extracted from source-free regions and examined for periods of high backgrounds due to flares. No significant flares were detected and the cleaned exposure times for the pointings are listed in Table~\ref{xray_obs}.

For both imaging and spectral analyses, we use the period E blank sky observations\footnote{See cxc.harvard.edu/contrib/maxim/acisbg}. The blank sky files were processed in the same manner as the observations and reprojected onto the sky to match the observations. For imaging and spectral analysis, the backgrounds are normalized to match the observation counts in the 10--12\,keV band where the \chan\ effective area is close to zero and the counts are dominated by the particle background. This procedure leads to background subtraction accurate to a few percent for energies $> 2\,$keV. However, the softer, diffuse X-ray background is known to vary over the sky. To account for this, we extract spectra from the S1 chip which is not contaminated by point sources and is free from cluster emission. There is a clear residual excess below 2\,keV after background subtraction. This excess is well-fitted by two unabsorbed MEKAL models with abundance set to the solar value and temperatures of kT=$0.20\pm0.01\,$keV and kT=$0.82\pm0.03\,$keV. Including this soft component, scaled by the ratio of the region areas, in the spectral fits performed in Section~\ref{temps} result in a $\sim 7\%$ increase in the measured temperatures.

\section{Cluster kinematics and substructure}\label{optical}

\subsection{Cluster Membership}\label{sec:memsel}

Cluster membership was achieved using a two-step approach. First, we identify candidate cluster members as those galaxies lying within a projected radial distance of 3.5\,Mpc from the brightest cluster galaxy (R.A.=14:15:08.39, Decl.=$-$00:29:35.7), having a redshift quality $nQ \ge 3$ and a peculiar velocity (defined with respect to the GAMA redshift of the BCG, $z_{BCG}=0.1389$) of $c(z-z_{BCG})/(1+z_{BCG}) = \pm 5000\,$\kms. This serves as a coarse first cut membership allocation and 481 galaxies are selected in this first step. The membership allocation is refined using the redshift-space distribution of galaxies and { iteratively} applying the caustics method of \citet[][]{diaferio1999}. The caustic amplitude is proportional to the escape velocity of the cluster and, therefore, determining the caustic amplitudes as a function of radius provides an excellent boundary with which cluster membership can be defined. Briefly, the distribution of galaxies in peculiar velocity-radius space is smoothed by a Gaussian kernel with an adaptive smoothing width (with $\sigma_v \neq \sigma_r$ where $\sigma_v$ and $\sigma_r$ are the smoothing widths in velocity and radius, respectively) which is proportional to the local density. { The local density is determined from the velocity-radius distribution which has been smoothed with a kernel of constant width, although again the smoothing width used for the velocity and radius are different. The constant smoothing widths are $\sigma_{r,const}= \sigma_{r, dist}/N^{1/6}$ and $\sigma_{v,const}= \sigma_{v, dist}/N^{1/6}$ where $\sigma_{r, dist}$ and $\sigma_{v, dist}$ are outlier-trimmed estimates of the standard deviations of the radial and peculiar velocity distributions, respectively. Here, $N$ is the number of galaxies assigned as cluster members. We then follow the basic method outlined in \citet{diaferio1999} to locate the caustic amplitudes and define membership based on the position of these caustics. We iterate the procedure until $N$ is stable and allow galaxies previously rejected as non-members to be reassigned as members if they fall within the latest iteration of the caustic boundaries. The caustics method has been shown to be an accurate mass estimator at large clustercentric radii \citep{rines2003,rines2006,serra2011,rines2012,alpaslan2012} and a robust method for allocation of cluster membership \citep{serra2013}. For further details, we refer the interested reader to the excellent explanations of the caustics method contained within the previously cited works.} 

\begin{figure}
\centering
\includegraphics[angle=0,width=.5\textwidth]{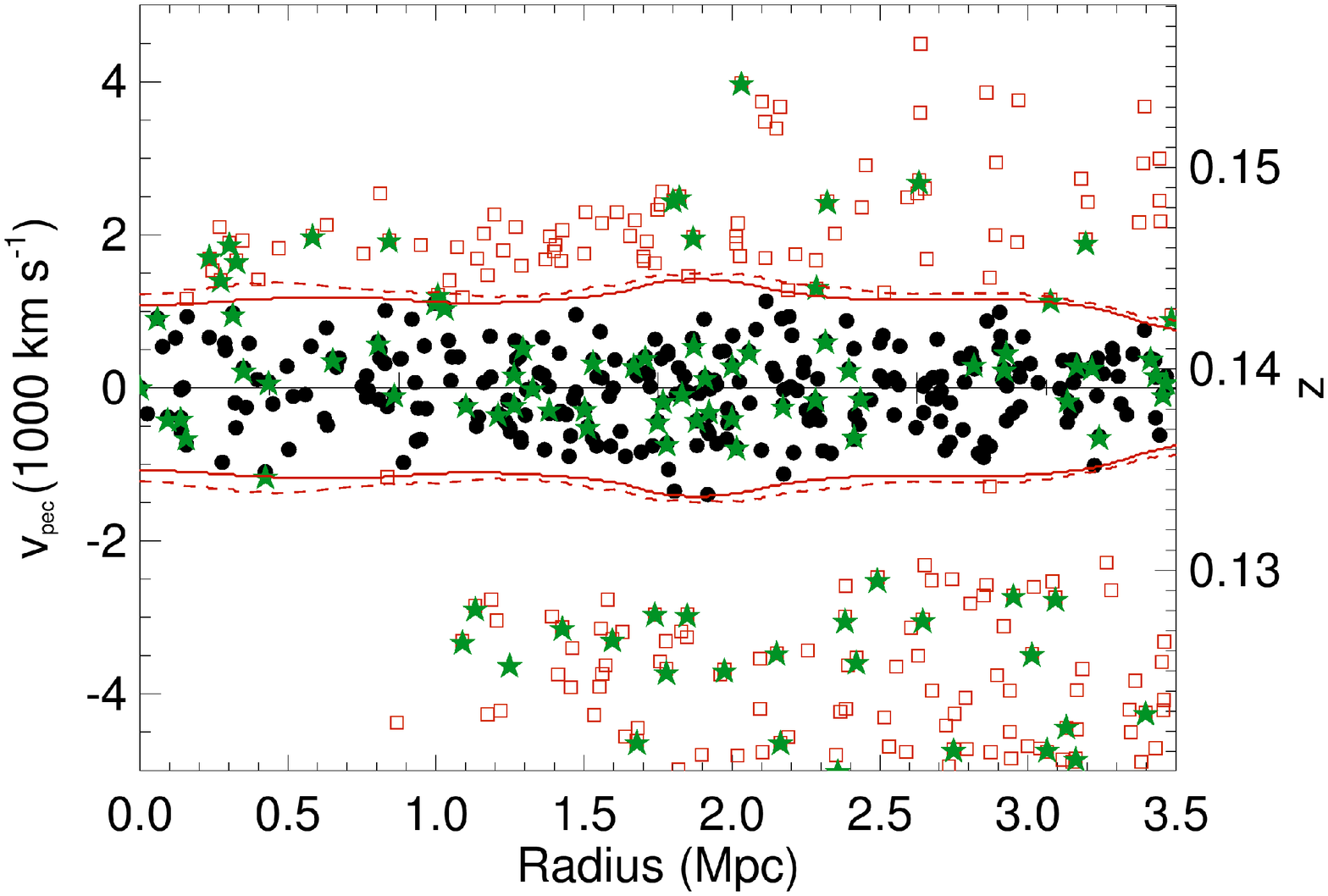}\\
\includegraphics[angle=0,width=.45\textwidth]{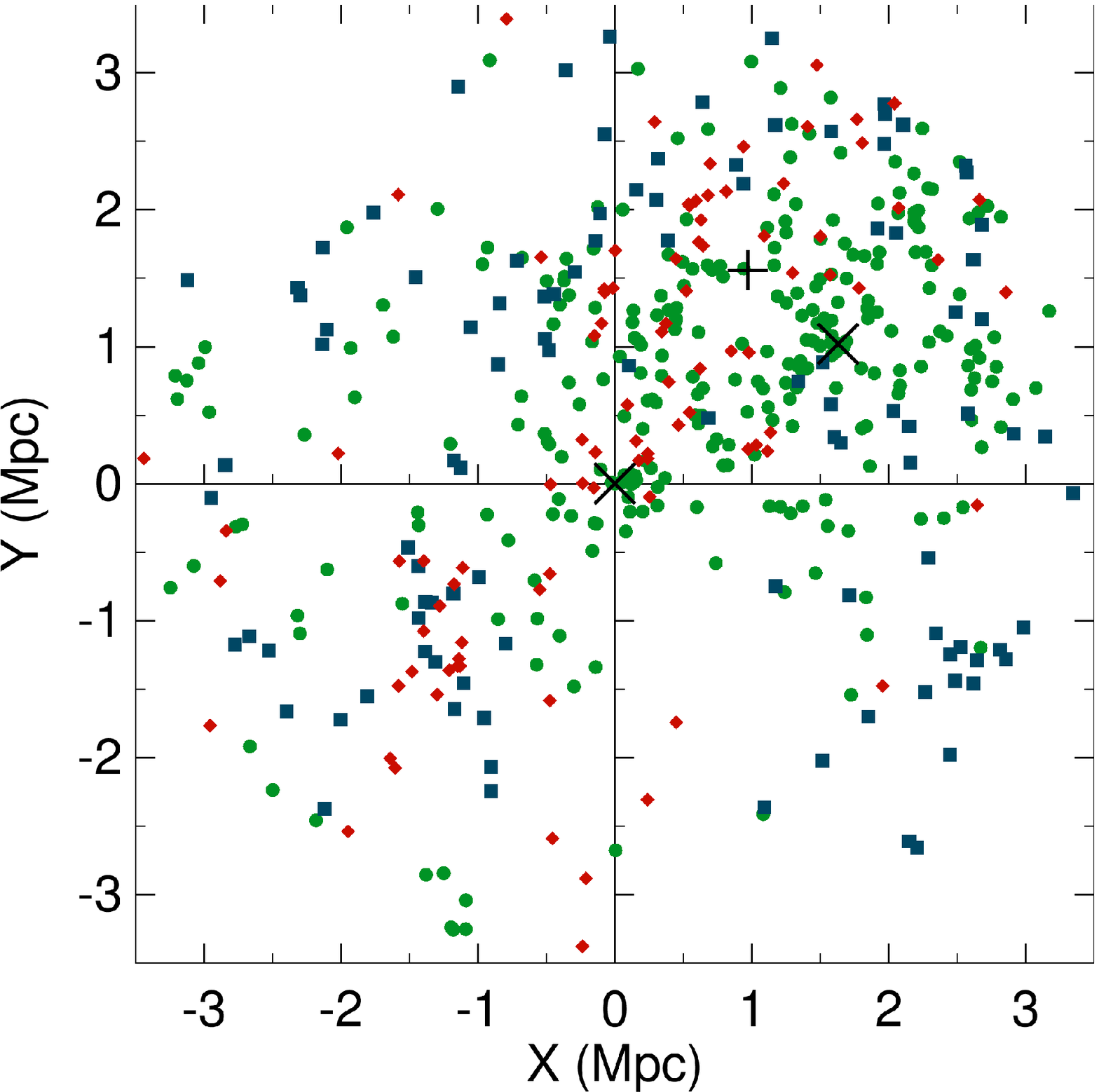}
\caption{Top panel: Phase-space diagram for galaxies with clustercentric distance within  3.5Mpc and $v_{pec} = \pm 5000\,$\kms\, of the BCG. The boundaries used to define cluster membership using the caustics method are shown with solid red lines. The outer dashed lines show the $1\sigma$ uncertainties on the caustic boundary (the inner uncertainty is not plotted). Filled black points show galaxies allocated as cluster members, open red squares show non-members and filled green stars show the distribution of SDSS redshifts. Note the significant enhancement in the source density due to GAMA's superior depth and completeness compared with the SDSS.
Bottom panel: Distribution of foreground (filled blue squares), member (filled green circles) and background (filled red diamonds) galaxies as determined from the caustics membership allocation. The spatial distribution of fore- and background galaxies is not strongly correlated with the spatial distribution of member galaxies indicating that our interloper rejection is robust. 
\label{vpec_rad}}
\end{figure}

The phase-space distribution is shown in the top panel of Figure~\ref{vpec_rad} where it can be seen that there is significant structure at $v_{pec}\simeq 2000\,$\kms\ which, given the offset at all radii from the main cluster body, is likely a background structure lying in projection along the line of sight (LOS). This structure makes locating the caustic amplitude difficult and for this reason we choose to use the well-separated negative $v_{pec}$ caustic amplitude to define the cluster membership. The caustic amplitude used to define cluster membership is shown in red, along with its associated uncertainty (shown only on the outer-side) which is determined as described in \citet{diaferio1999}. The final sample of spectroscopically confirmed cluster members contains 283 galaxies within a cluster-centric radius of 3.5\,Mpc. 

In the bottom panel of Figure~\ref{vpec_rad} we show the spatial distribution of the fore- and background galaxies within $c(z-z_{BCG})/(1+z_{BCG}) = \pm 5000\,$\kms as filled blue squares and filled red diamonds, respectively, along with the allocated members (filled green circles). The spatial distribution of the fore- and background galaxies is different from the distribution of member galaxies. This provides additional support for them not being associated with the cluster. We also show the positions of the two brightest cluster galaxies as crosses, along with the cluster center assigned to A1882 in \citet{abell1989} (R.A.=14:14:42, Decl.=-00:19:00) as a plus sign. We note that the Abell center is some $1835\,$kpc northwest of our assigned cluster center.

\subsection{The peculiar velocity distribution}\label{vpec_dist}

With our sample of cluster members, we use the biweight estimators \citep{beers1990} to measure a cluster redshift of $z_{clus}=0.1389\pm0.0002$, consistent with { the SDSS DR9 redshift measurement for the BCG of $z_{BCG}=0.13893\pm0.00003$ \citep{ahn2012}}. Also measured is the cluster velocity dispersion $\sigma_v = 525\pm23\,$\kms. The $1\sigma$ uncertainties on our cluster redshift and velocity dispersion measurements are determined with the jackknife resampling technique. It is worth noting that our redshift and dispersion measurements are significantly different from those values reported based on the C4 clustering algorithm \citep[$z=0.1404$, $\sigma_v=931\,$\kms\ from 48 members,][]{miller2005}, the RASS-SDSS cluster survey \citep[$z=0.1396$, $\sigma_v=733\,$\kms\ from 55 members,][]{popesso2007} and the GAMA Galaxy Group Catalog \citep[$z=0.1394$, $\sigma_v=833\,$\kms\ from 264 members,][]{robotham2011}. { We can reproduce the results of these earlier works by including galaxies with $c|(z-z_{BCG})|/(1+z_{BCG}) \leq 2600\,$\kms\ as cluster members. This selection includes the background interlopers removed by our caustics member allocation. Remeasuring the biweight estimators for this modified member selection, we find $z_{clus, mod}=0.1402\pm0.0002$ and $\sigma_{v, mod} = 932\pm50\,$\kms, consistent with the larger redshift and velocity dispersion from earlier results.} This indicates that these studies were affected by interloper contamination from the background galaxies that our member selection technique successfully rejected. Indeed, the ``modality'' and kurtosis measurements provided in the \citet{robotham2011} catalogs indicate significant departures from a Gaussian shape, likely due to the effect of interlopers.

\begin{figure}
\centering
\includegraphics[angle=0,width=.45\textwidth]{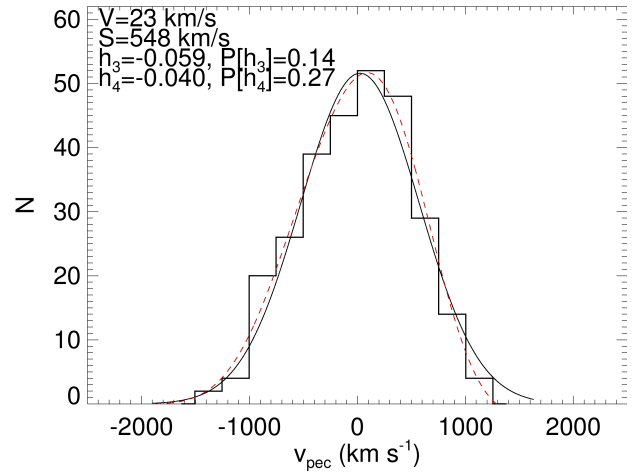}
\caption{The histogram shows the distribution of peculiar velocities for member galaxies within 3.5Mpc of the BCG in A1882. The solid black line shows the best fitting Gaussian with parameters shown in the upper left of the plot. The dashed red line shows the Gauss-Hermite reconstruction of the observed distribution. The $h_3$ and $h_4$ terms, representing symmetric and asymmetric deviations from a Gaussian shape, respectively, are not significant and indicate that the shape of the observed distribution does not differ significantly from a Gaussian.
\label{vpec}}
\end{figure}

For a dynamically relaxed cluster, the peculiar velocity distribution is well approximated by a Gaussian shape. The large peculiar velocities induced during cluster mergers can perturb this Gaussian shape, particularly when the merger is close to core-passage and the merger axis is aligned close to our LOS \citep[e.g., A2744,][]{owers2011a}. These perturbations are detected as higher order moments in the velocity distribution, such as the skewness and kurtosis. Here we use the Gauss-Hermite reconstruction technique to test for non-Gaussianity in the peculiar velocity distribution \citep[see][for a detailed description]{zabludoff1993,owers2009a}. Briefly, the velocity distribution is described by a series of Gauss-Hermite functions with the Gauss-Hermite moments $h_0\simeq1$ multiplying the best-fitting Gaussian with mean $V$ and dispersion $S$, while the $h_3$ and $h_4$ terms describe asymmetric and symmetric deviations from a Gaussian shape, respectively. The peculiar velocity distribution is shown in Figure~\ref{vpec}, along with the best-fitting Gaussian and Gauss-Hermite reconstructions (solid black, and dashed red curve, respectively). The distribution is well described by a Gaussian with the measured values of $h_3=-0.059$ (implying a negative skewness at the $6\%$ level) and $h_4=-0.040$ (implying the distribution is broader than a Gaussian at the $4\%$ level) occurring in $\sim 15\%$ and $\sim 27\%$, respectively, of 5000 simulations of Gaussian random distribution with the same number, mean and dispersion as the data.

\subsection{Spatial distribution of Member Galaxies}\label{adapt_smooth}

While the peculiar velocity distribution is an excellent probe for detecting high velocity merger aligned along our LOS, it is generally a poor indicator for mergers which are occurring with the majority of their motion directed perpendicular to our LOS, e.g., the well-known major merger Abell~3667 has a velocity distribution which is well described by a single Gaussian distribution \citep{owers2009b}. Here, the spatial distribution of the member galaxies are an excellent probe of substructure, particularly when the merging structures are well separated and, therefore, easily discernible as enhancements in the surface density of the galaxies. The isopleths shown in \citet{morrison2003} reveal complex multimodality in the spatial distribution of galaxies in the direction of A1882. However, those isopleths are generated without the aid of spectroscopic redshifts and may be significantly affected by contamination from unassociated structure lying along the LOS toward A1882. We have shown in Section~\ref{sec:memsel} that there exists a great deal of background structure lying in the direction of A1882. Using our spectroscopically confirmed members defined in Section~\ref{sec:memsel}, we can asses whether the rich structure seen by \cite{morrison2003} is present in our sample. 

We have applied a two-step adaptive smoothing algorithm to the spatial distribution of the cluster members in order to reveal local overdensities in the galaxy surface density. On the first pass the spatial distribution is smoothed by a Gaussian kernel with an optimum width which is proportional to number of cluster members, $N$, and the standard deviation of the spatial distribution, $\sigma_x$ and $\sigma_y$, such that $\sigma_{opt} = 0.96\sqrt{0.5*(\sigma_x + \sigma_y)}*N^{-1/6}$ \citep{silverman1986}. This provides an initial estimate of the galaxy surface density distribution. The initial density estimate is used in the second pass to define an adaptive smoothing kernel with width $\sigma^{adapt}_{x,y}=\lambda_{x,y} \sigma_{opt}$ where $\lambda_{x,y} = (g/\Sigma^{init}_{x,y})^{1/2}$, $g$ is the geometric mean of the first-pass density distribution and $\Sigma^{init}_{x,y}$ is the initial estimate of the density at position of interest. The results of this smoothing procedure are shown as black contours in Figure~\ref{bubbles}. Consistent with the analysis of \citet{morrison2003}, the galaxy surface density shows a rich array of local peaks. There are two prominent peaks; one is associated with the BCG (within 100\,kpc) and the second, located $\sim 2\,$Mpc to the northwest, lies within 100\,kpc of the second rank cluster galaxy.

\subsection{Localized kinematical substructure}

Having identified the existence of multiple local peaks in the spatial distribution of galaxies, we now wish to determine if these structures are also kinematically distinct. This is achieved by using the $\kappa$-test \citep{colless1996} to search for departures of the local kinematics around each galaxy in the member sample from the global cluster kinematics. We define ``local'' as the $n_{loc}=\sqrt{N_{mem}}$ nearest neighbors to the galaxy of interest where $N_{mem}$ is the number of cluster members in our sample. The peculiar velocity distribution of the near neighbors is compared to the global velocity distribution using the Kolmogorov-Smirnov test to assess the likelihood, $P_{KS, i}$, that the local and global distributions are drawn from the same parent distribution. { We note that in defining the global velocity distribution, we have excluded the $n_{loc}=\sqrt{N_{mem}}$ members local to the galaxy of interest.} A measure of the overall kinematical substructure present within the cluster is determined by the summation $\kappa_{tot}=\sum_{i=0}^{N_{mem}} -log P_{KS, i}$. The significance of $\kappa_{tot}$ is determined by comparison to the distribution of 10,000 Monte Carlo realizations of $\kappa_{ran}$. These realizations are produced by fixing the spatial coordinates for each cluster member and randomly shuffling the peculiar velocities, thereby erasing any correlation between position and velocity, and measuring $\kappa_{ran}$. The $\kappa_{tot}=214$ value lies $\sim 3.8\sigma$ from the mean of the distribution of the $\kappa_{ran}$ values and $\kappa_{ran} \geq \kappa_{tot}$ does not occur in the 10,000 realizations. Thus, there is only a very small probability that the observed $\kappa_{tot}$ value has occurred by chance.

\begin{figure*}
\centering
\includegraphics[angle=0,width=.45\textwidth]{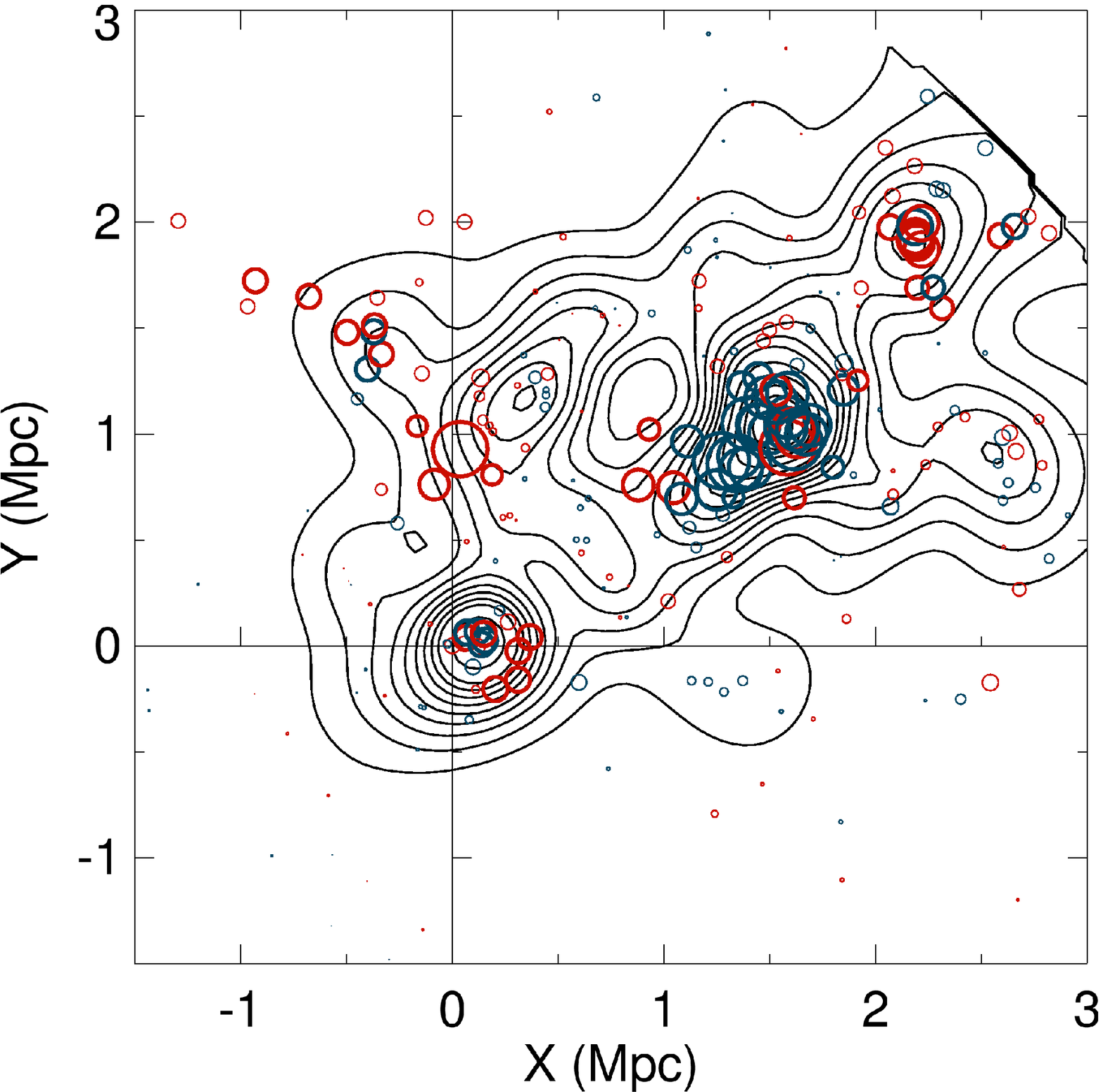}
\includegraphics[angle=0,width=.45\textwidth]{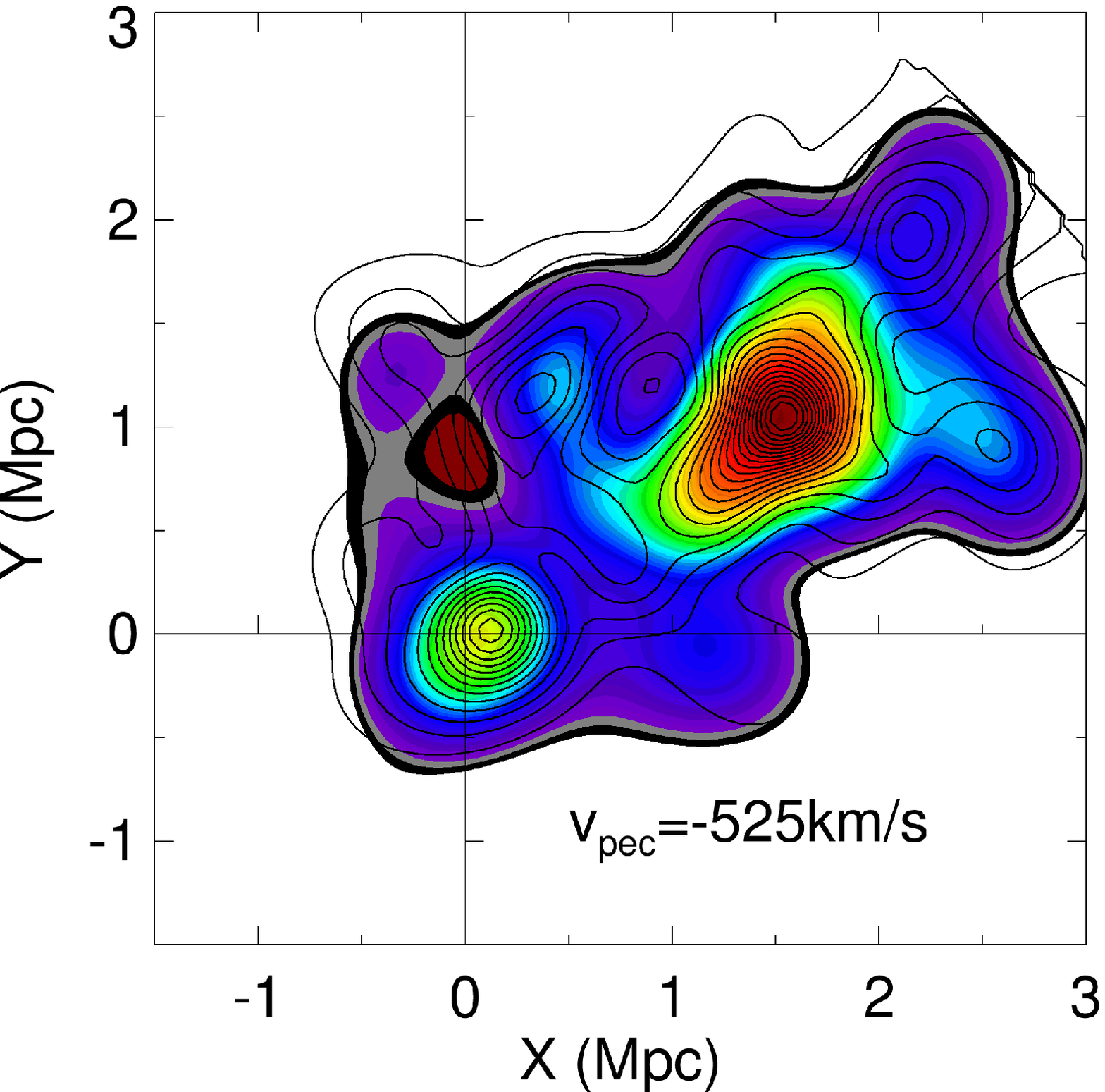}\\
\includegraphics[angle=0,width=.45\textwidth]{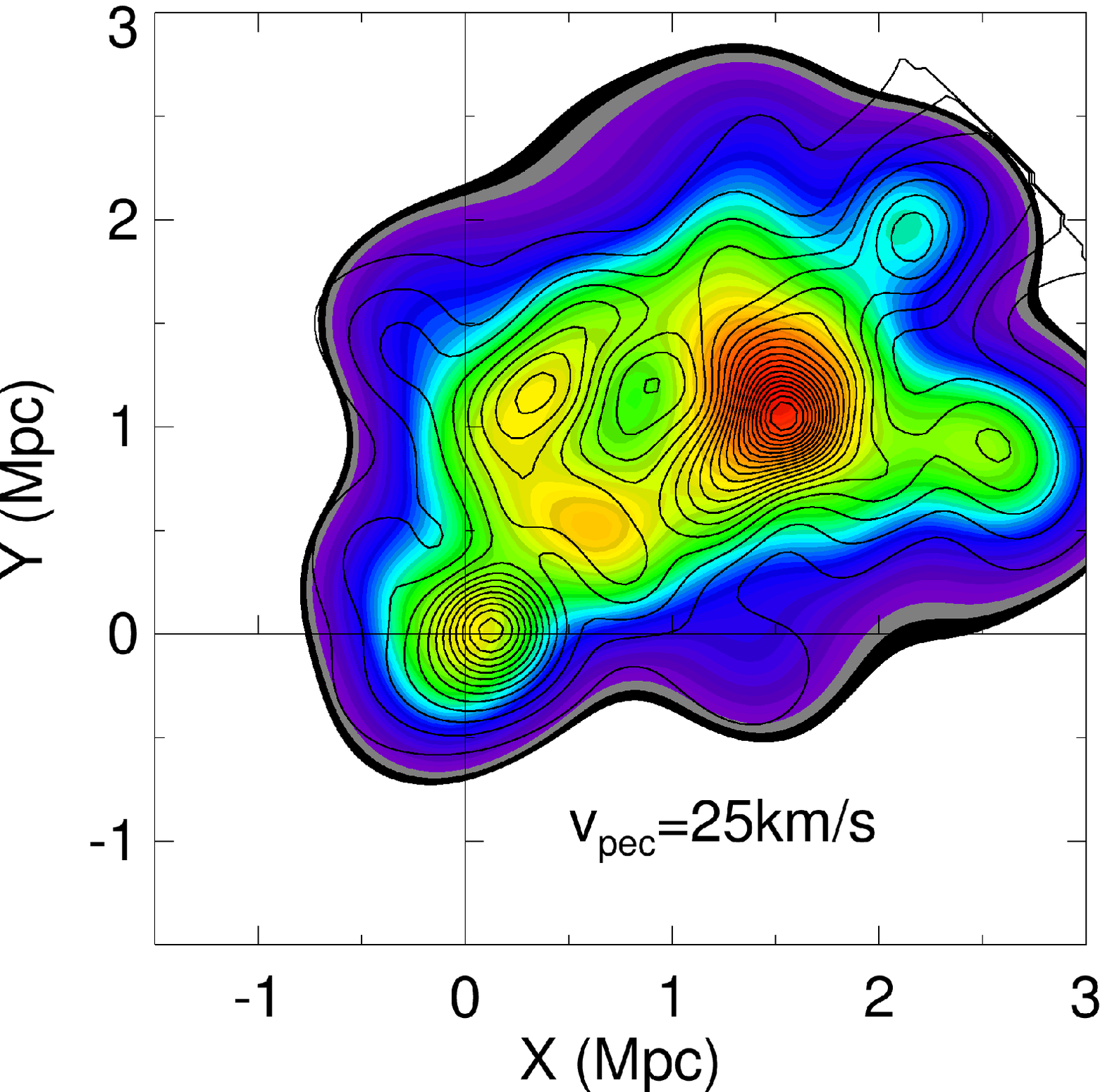}
\includegraphics[angle=0,width=.45\textwidth]{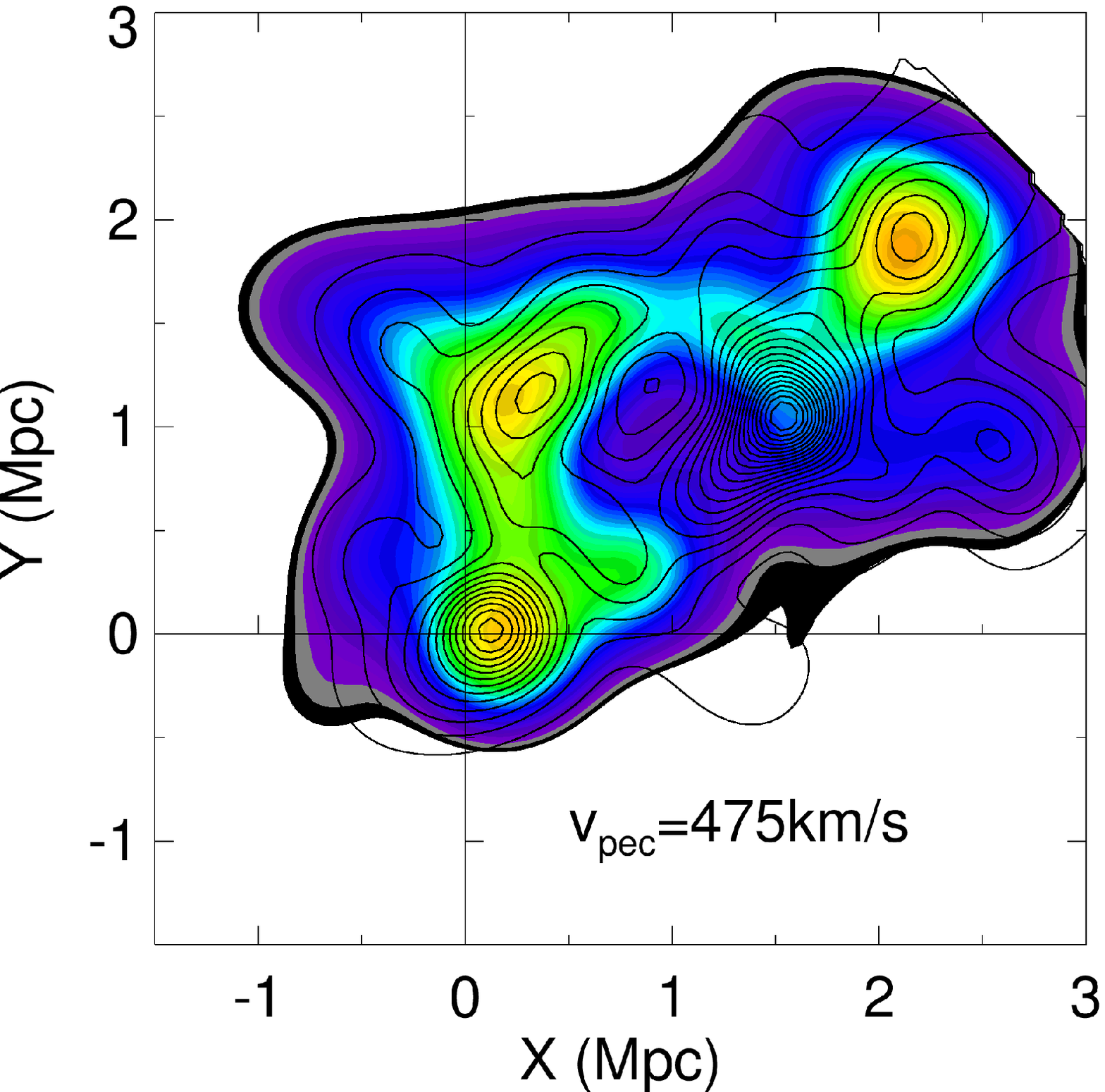}\\

\caption{Top left panel: ``Bubble'' plot where the circle radius is proportional to $-log P_{KS, i}$ where $P_{KS,i}$ is the KS probability that the local and global peculiar velocity distribution are drawn from the same parent distribution. Clusters of large, emboldened circles indicate significant local departures from the global kinematics. The circles are color coded so that galaxies with negative and positive peculiar velocities are blue and red, respectively. The top right and lower panels show the three ``tomograms'' which reveal where the color map shows the 3D density at three central velocities; -525\kms, 25\kms\, and 475\kms, with red colors indicating the highest density and purple colors low density.
The black contours in each panel show the adaptively smoothed distribution of member galaxies. The contour levels range from 10-100 galaxies ${\rm Mpc}^{-2}$ and the contours increment by 5 and 10 for the levels 10-50 and 50-100, respectively.
\label{bubbles}}
\end{figure*}

The results of the $\kappa$-test are best visualized in the form of ``bubble'' plots where, at the position of each member galaxy, a circle with radius $r \propto -log P_{KS, i}$ is plotted. Kinematical substructures are revealed by regions containing clusters of large circles in Figure~\ref{bubbles}. Where the departure is significant, i.e., the value of $P_{KS, i}$ occurs in only $5\%$ of the Monte Carlo realizations, we plot an emboldened circle. To give an indication as to whether the departure in the local kinematics is due to a local deviation in the peculiar velocity, we have colored the circles red or blue in order to indicate positive or negative peculiar velocities with respect to the cluster mean, respectively. A number of significantly large, blue bubbles are associated with the local peak in the galaxy density distribution located $\sim2\,$Mpc to the northwest, strongly indicating that it is also distinguished as a local kinematical substructure with a negative peculiar velocity. The bubble plot shows a number of significantly large circles near to the BCG, indicating evidence for kinematical substructure there. { However, the colors of the circles indicate that there is no preference for either negative or positive peculiar velocities in this region. Therefore, the significant departure of the local kinematics from the global kinematics revealed here is likely to be due to differences in the shapes of the local and global velocity distributions, rather than due to a large local peculiar velocity difference. For example, the velocity dispersion may be enhanced in this region compared with the global velocity distribution. Alternatively, the contribution of other localized kinematical substructures to the global velocity distributions may lead to differences between the local and global velocity distributions.} Finally, the less-significant local peak in the galaxy surface density distribution which is $\sim 3\,$Mpc northwest of the BCG also harbors significant kinematical substructure and the galaxies have systematically higher peculiar velocities in this region.

The majority of the kinematical substructures, particularly the NW one, show a preference for having either positive or negative peculiar velocities. A better indication of where these structures lie in peculiar velocity space can be gleaned from the 3D smoothed galaxy density distribution. For the spatial portion of the smoothing, we apply the same adaptive kernel as outlined in Section~\ref{adapt_smooth} while in the velocity direction we smooth with a Gaussian kernel with a constant width of $\sigma=250\,$\kms. In the top right and bottom panels of Figure~\ref{bubbles} we present tomograms showing the smoothed 3D galaxy density in velocity slices centered at $v_{pec}=-525, 25\, {\rm and}\, 475\,$\kms. These velocity positions were chosen from the full velocity range because they reveal where peaks in the 3D distribution lie in velocity space.

\subsection{Characterising the substructure}\label{kmm}

The analyses presented above reveal a significant substructure $\sim 2\,$Mpc\ to the northwest of the main cluster in A1882 which is both spatially and kinematically distinct. Hereafter, we label this northwestern substructure A1882B, while the central, main cluster is labeled A1882A. We now wish to more accurately constrain the kinematics of A1882B with the aim of obtaining a better understanding of the mass and merger history of this system. To do this, we utilize the Kaye's Mixture Modeling algorithm \citep{ashman1994} to estimate the mean velocities and dispersions of A1882A and A1882B. This algorithm has been extensively used for characterizing substructure within clusters \citep[e.g.,][]{colless1996,barrena2002,boschin2006,maurogordato2008,girardi2008,owers2009a,owers2009b,owers2011a}. 

The algorithm fits a user-specified number of N-dimensional Gaussians to the data using the maximum likelihood method to determine the best fitting parameters. In our case, we wish to exploit all of the available data, and so we fit the full 3D distribution of galaxies. While the assumption of a Gaussian shape for the spatial distribution of galaxies in a cluster is not physically well motivated the KMM methodology has been shown to work well with spatial information alone \citep{kriessler1997}. Another limitation of the KMM algorithm is that it can be significantly affected by the presence of outliers. Preliminary test runs of the algorithm indicate that the two substructures located to the northwest at radii $>2.5\,$Mpc significantly affect the stability of the KMM fits to A1882B. Thus, we exclude galaxies beyond 2.5\,Mpc for the remainder of our analysis. 

\begin{figure}
\centering
\includegraphics[angle=0,width=.5\textwidth]{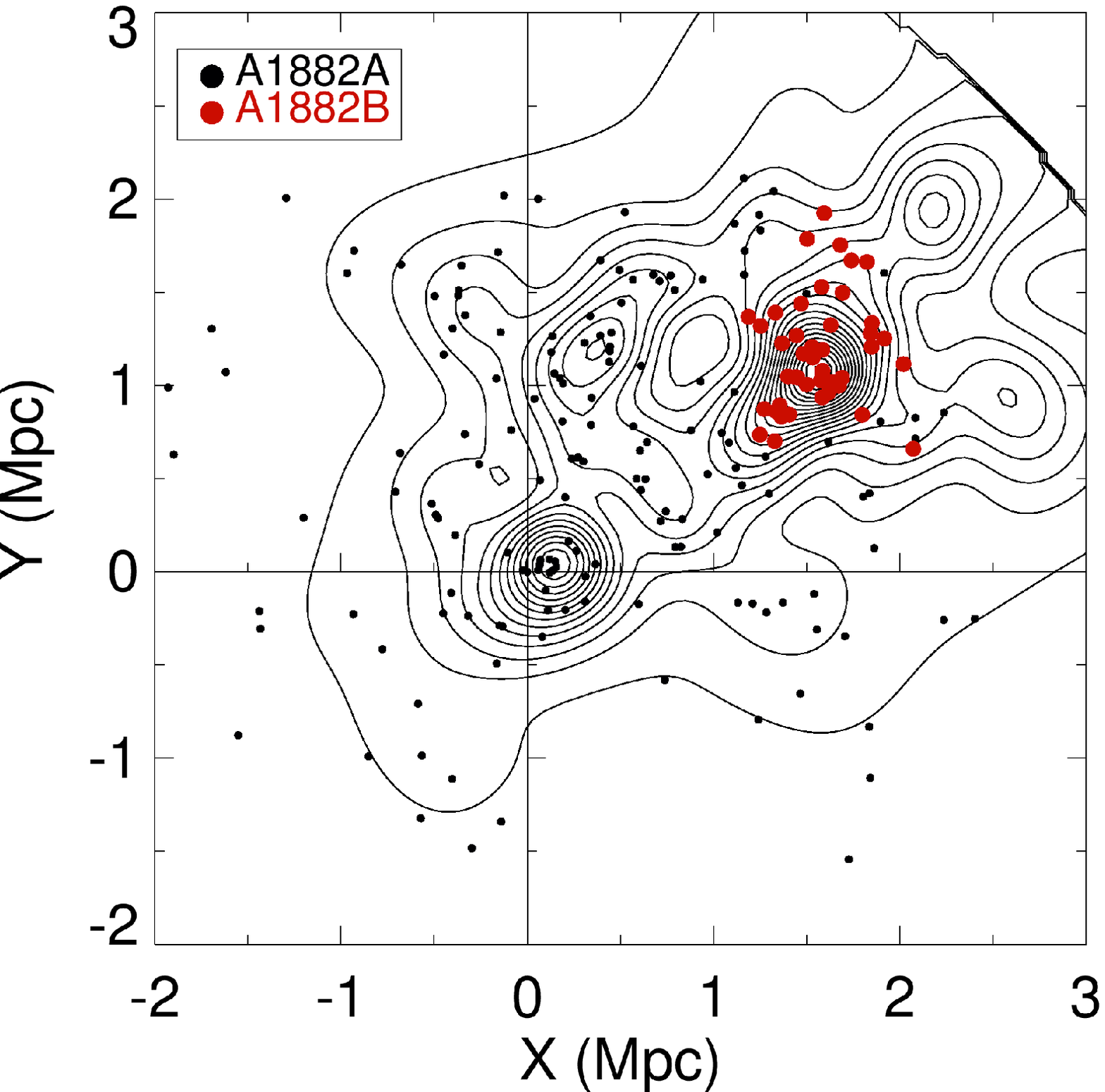}\\
\includegraphics[angle=0,width=.5\textwidth]{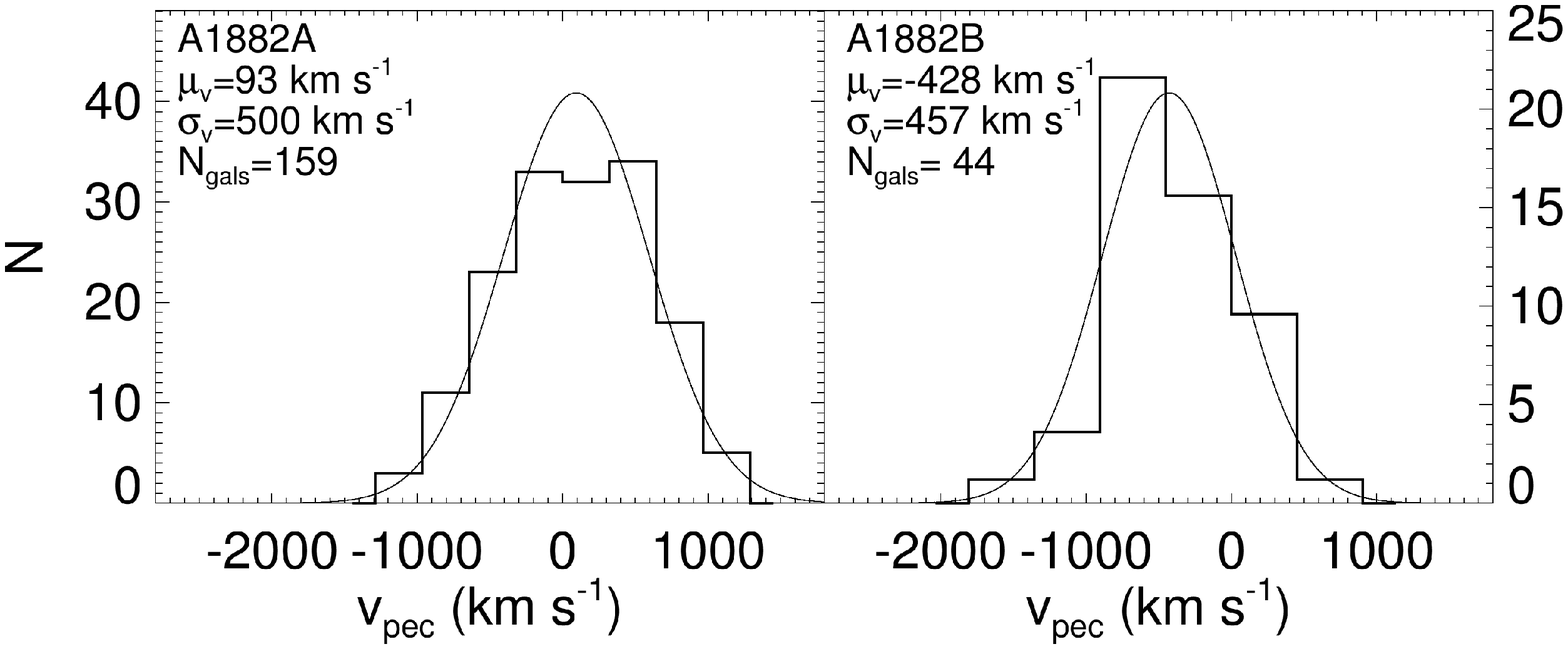}
\caption{Top panel: The spatial distribution of galaxies allocated by KMM  to the primary structure, A1882A, are shown as filled black circles, while those allocated to the secondary, A1882B, are shown as large red filled circles. The black contours show the same adaptively smoothed galaxy surface density distribution as in Figure~\ref{bubbles}. Lower panel: The $v_{pec}$ distribution of galaxies allocated to A1882A (left) and A1882B (right) by the KMM algorithm. The relevant details of the fit parameters are presented in Table~\ref{kmm_fits}.
\label{kmm_imgs}}
\end{figure}

The KMM algorithm requires fairly robust initial estimates of the 3D positions and dispersions of the substructures, as well as estimates of the fraction of galaxies within the substructure. The initial estimates for the spatial positions of the substructures are obtained from the peaks in the galaxy surface density distribution Section~\ref{adapt_smooth}. The initial mean velocities and velocity dispersions are estimated in apertures surrounding the local peaks in the galaxy surface density distribution and are listed in Table~\ref{kmm_fits} along with the parameters returned by the best fitting KMM models. Uncertainties on these parameters are determined from the distribution of 5000 non-parametric bootstrap resamplings of the data where KMM has been re-run on the resampled data, producing new best-fitting parameters. The top panel in Figure~\ref{kmm_imgs} shows the spatial distribution of the galaxies allocated to the A1882A and A1882B by the KMM algorithm, as well as the corresponding velocity histograms (lower panel, Figure~\ref{kmm_imgs}).

In its original form the KMM algorithm was designed to assess one-dimensional distributions for bimodality and return a P-value which provides a quantitative assessment of the improvement in going from a unimodal to a bimodal fit. This is achieved by comparing the likelihood ratio test statistic (LRTS) to a chi-squared distribution. However, as noted in \citet{ashman1994} the P-value only gives a useful indication of the improvement in the fit for the specific case of a one-dimensional, homoscedastic (i.e., equal variances) bimodal versus unimodal fit. Our data fail these criteria on two accounts; they are three-dimensional and non-homoscedastic. We overcome this issue using the method described in \citet{owers2011a}, i.e., we use parametric bootstrapping to determine the probability of obtaining a LRTS as large as that observed. Briefly, this is achieved by resampling the best fitting single Gaussian 3D model 5000 times, refitting for both the single and two-Gaussian cases using the same input estimates listed in Table~\ref{kmm_fits}, and determining the distribution of the LRTSs. As can be seen in Table~\ref{kmm_fits}, the P-value returned by this analysis is low (in fact, none of the bootstrap LRTSs were as high as the observed one), thus the two-mode fit provides a much better description of the data than does a one-mode fit.

\begin{deluxetable*}{ccccccc}
\tabletypesize{\scriptsize}
	\tablecolumns{7}
  \tablecaption{Results from the 3D KMM clustering analysis for two partitions ($N_g=2$) and excluding galaxies with radius $> 2.5\,$Mpc.\label{kmm_fits}}
\tablehead{   \colhead{Structure}     &     \multicolumn{2}{c}{Initial Input} & \multicolumn{2}{c}{Final Output} & \colhead{Nmem} &\colhead{P($N_g-1, N_g$)} \\%
            & \colhead{($\overline{x},\overline{y},\overline{v}$)}      & \colhead{($\sigma_x,\sigma_y, \sigma_v$)} &  \colhead{($\overline{x},\overline{y},\overline{v}$)}      & \colhead{($\sigma_x,\sigma_y, \sigma_v$)} & &}
\startdata
Primary  (A1882A)      & ($608, 676, 29$) & ($988, 801, 565$)     &  ($343^{+105}_{-131}, 538^{+80}_{-98}, 93^{+33}_{-57}$)  & ($942^{+54}_{-68}, 833^{+31}_{-57}, 500^{+23}_{-26}$) & 159 & -- \\
Secondary (A1882B)      &  ($1472, 1120, -566$) &($354, 422, 439$) &($1565^{+44}_{-94}, 1172_{-131}^{+107}, -428^{+187}_{-139}$) &($209^{+58}_{-37}, 299^{+93}_{-80}, 457^{+108}_{-101}$) & 44 &  0.00\\
\enddata
\end{deluxetable*}

\section{X-ray structure and temperature distributions}\label{xray}

Detection of the optically defined substructures at X-ray wavelengths will confirm their nature as significant substructures while their X-ray temperatures can be used to estimate their masses. Moreover, the collisional nature of the ICM means that past or ongoing merger activity may be revealed as features detected in X-ray images or in the derived thermodynamic maps \citep[e.g., shocks, cold fronts, multiple components][]{markevitch2002,markevitch2007, owers2009c, russell2010, owers2011a}. Therefore, the X-ray data provides an excellent diagnostic of the recent merger history which complements the kinematical information provided by the optical spectroscopy.

\subsection{Imaging analysis}

\begin{figure*}
\centering
\includegraphics[angle=0,width=.47\textwidth]{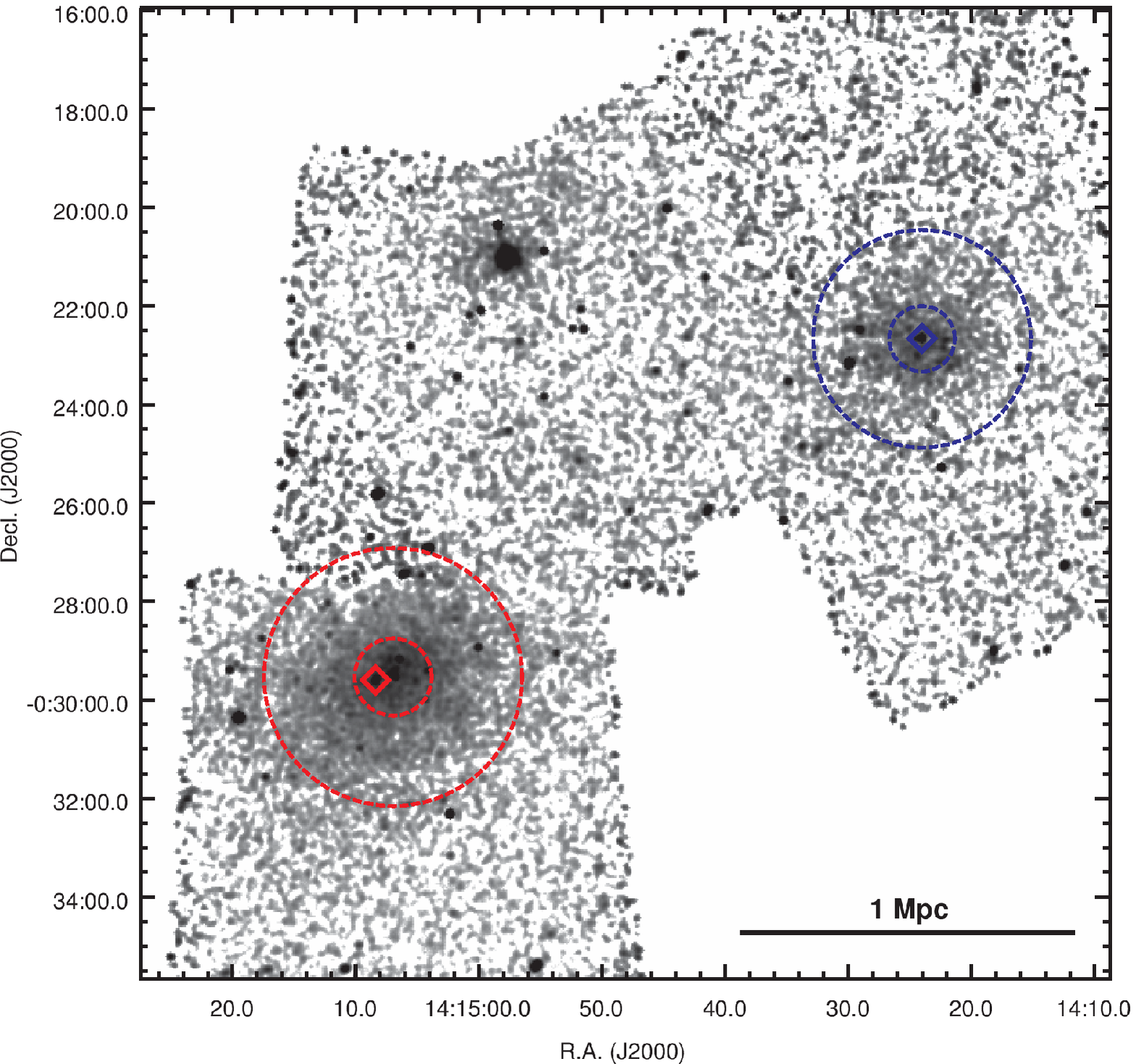}
\includegraphics[angle=0,width=.47\textwidth]{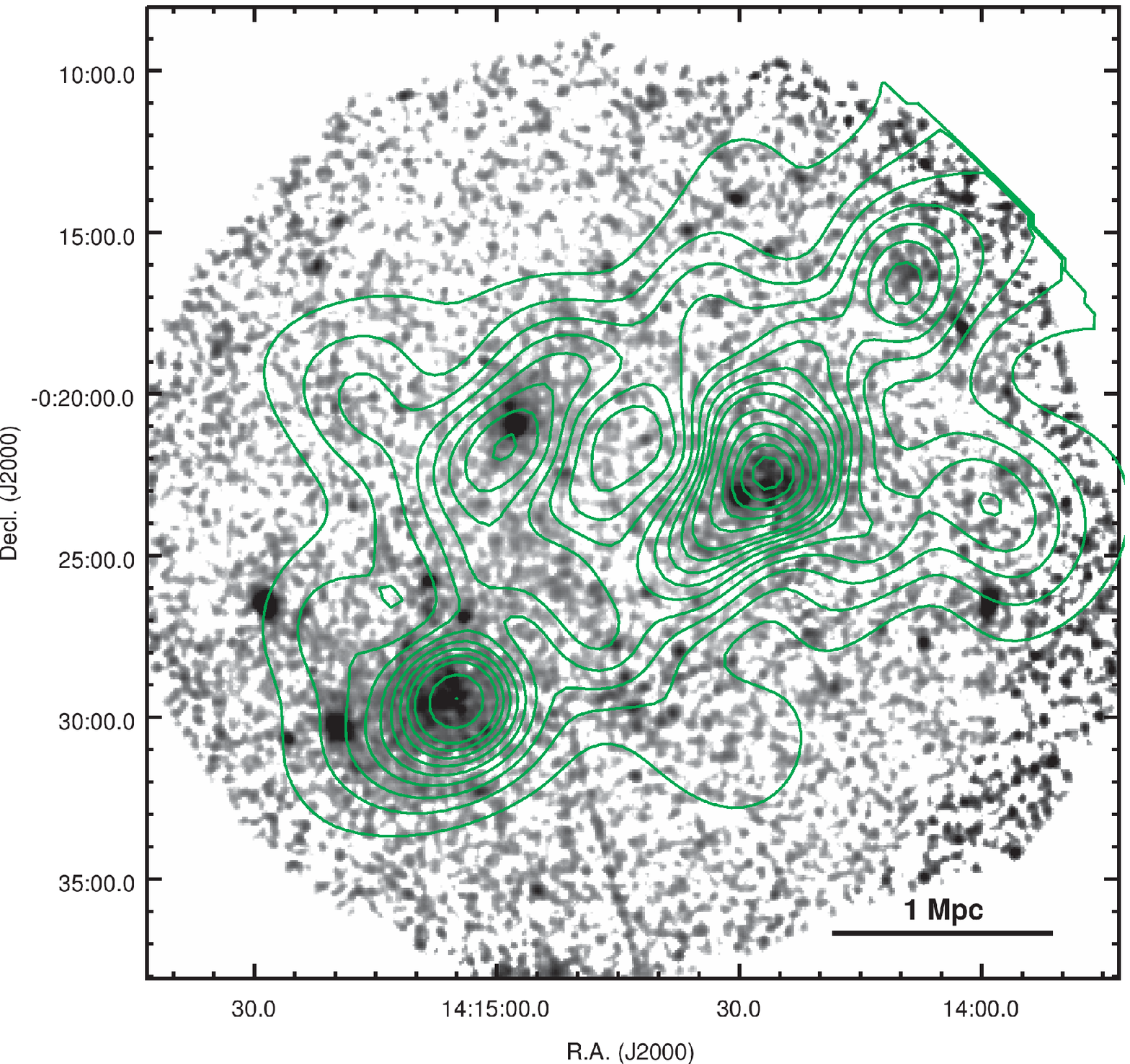}
\caption{Left panel: Combined, background subtracted and exposure corrected \chan\ image. A light ($6''$ FWHM) smoothing has been applied to the image. The red and blue diamonds show the positions of the BCGs associated with A1882A and A1882B, respectively. The red and blue annuli show the $0.15-0.5r_{500}$ regions used to extract the spectra to determine the subcluster temperatures in Section~\ref{temps}. { Right Panel: Combined, background subtracted and exposure corrected XMM image smoothed with a Gaussian kernel ($12''$ FWHM). The green contours show the same galaxy isopleths that are presented in Figure~\ref{bubbles}}.
\label{xray_imgs}}
\end{figure*}

{ The combined, background subtracted and exposure corrected \chan\ and XMM images are shown in the left and right panels of Figure~\ref{xray_imgs}, respectively. The green contours overlaid onto the XMM image show the same density contours as those presented in Figure~\ref{bubbles}. There is clearly diffuse, extended emission associated with both A1882A and A1882B confirming their nature as bona-fide, gravitationally bound systems.} The emission associated with A1882A and A1882B appears fairly regular and lacks obvious edges or structures associated with merger activity. We also note the dearth of any bright core emission associated with a cool core. The emission associated with A1882A appears elongated with position angle roughly aligned along a SE-NW axis. A number of point sources are spread across the field, many of which are associated with cluster members, and the most notable of which is located $\sim 8.8\,$\arcmin\ to the north of A1882A and is associated with a spectroscopically confirmed cluster member. These bright point sources are easily identifiable in the high-resolution, deep \chan\ images, but are likely to have gone undetected in the \citet{morrison2003} study, where the lower resolution, shallower {\it ROSAT} All Sky Survey images were used to measure the X-ray luminosity. This is likely to have lead to a poor (probably over) estimate of the X-ray luminosity for A1882.

\begin{figure*}
\centering
\includegraphics[angle=0,width=.45\textwidth]{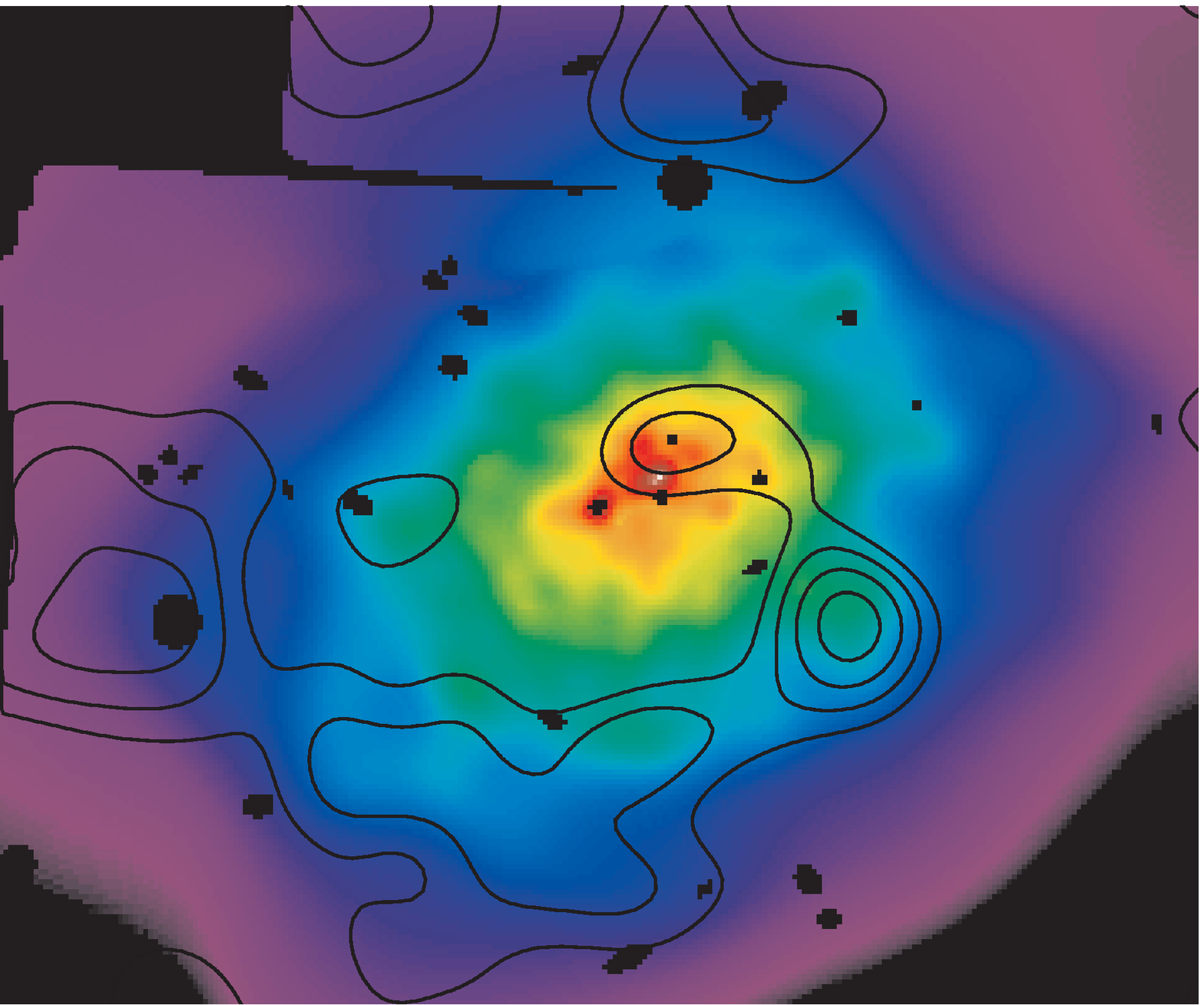}
\includegraphics[angle=0,width=.45\textwidth]{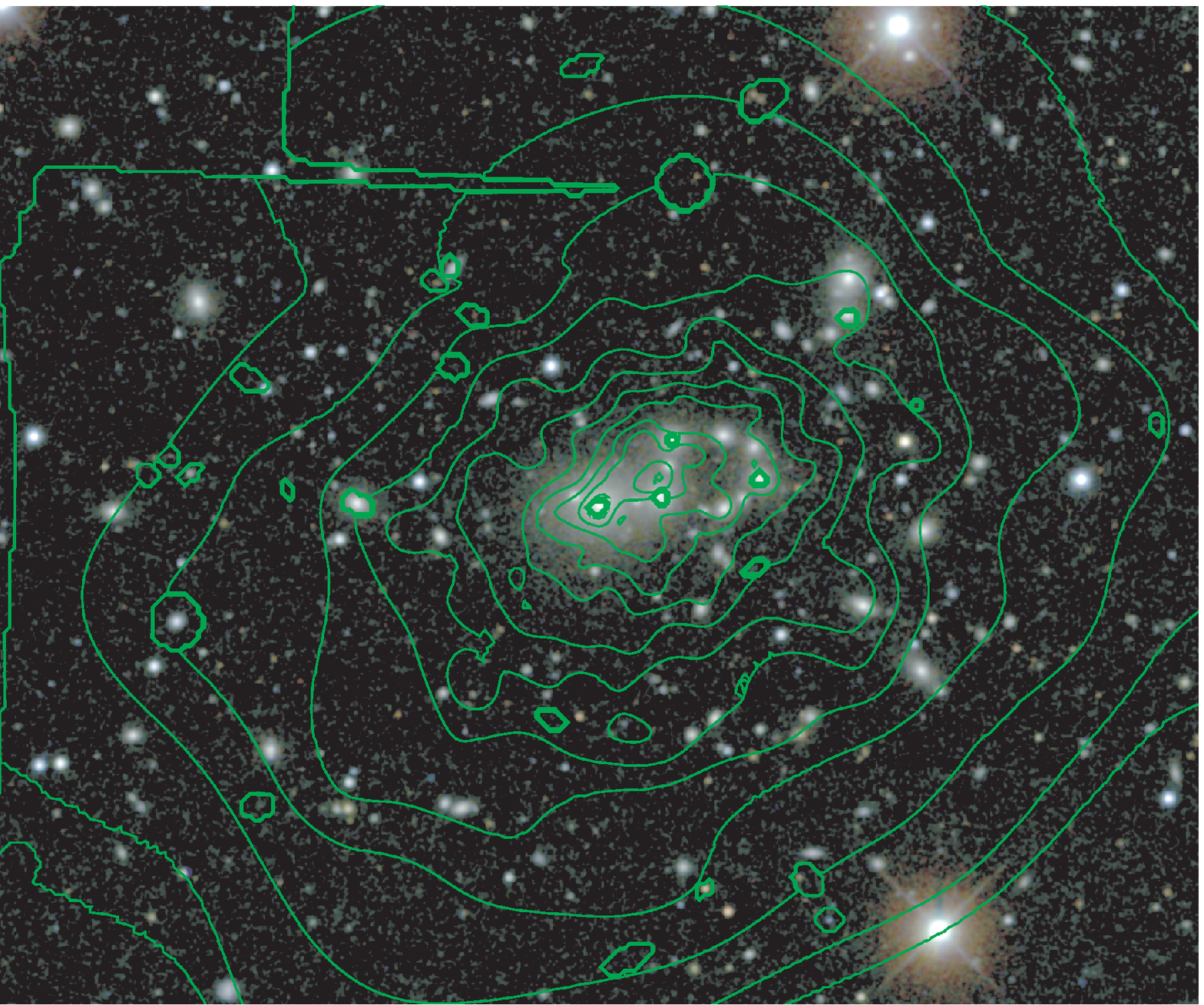}\\
\includegraphics[angle=0,width=.45\textwidth]{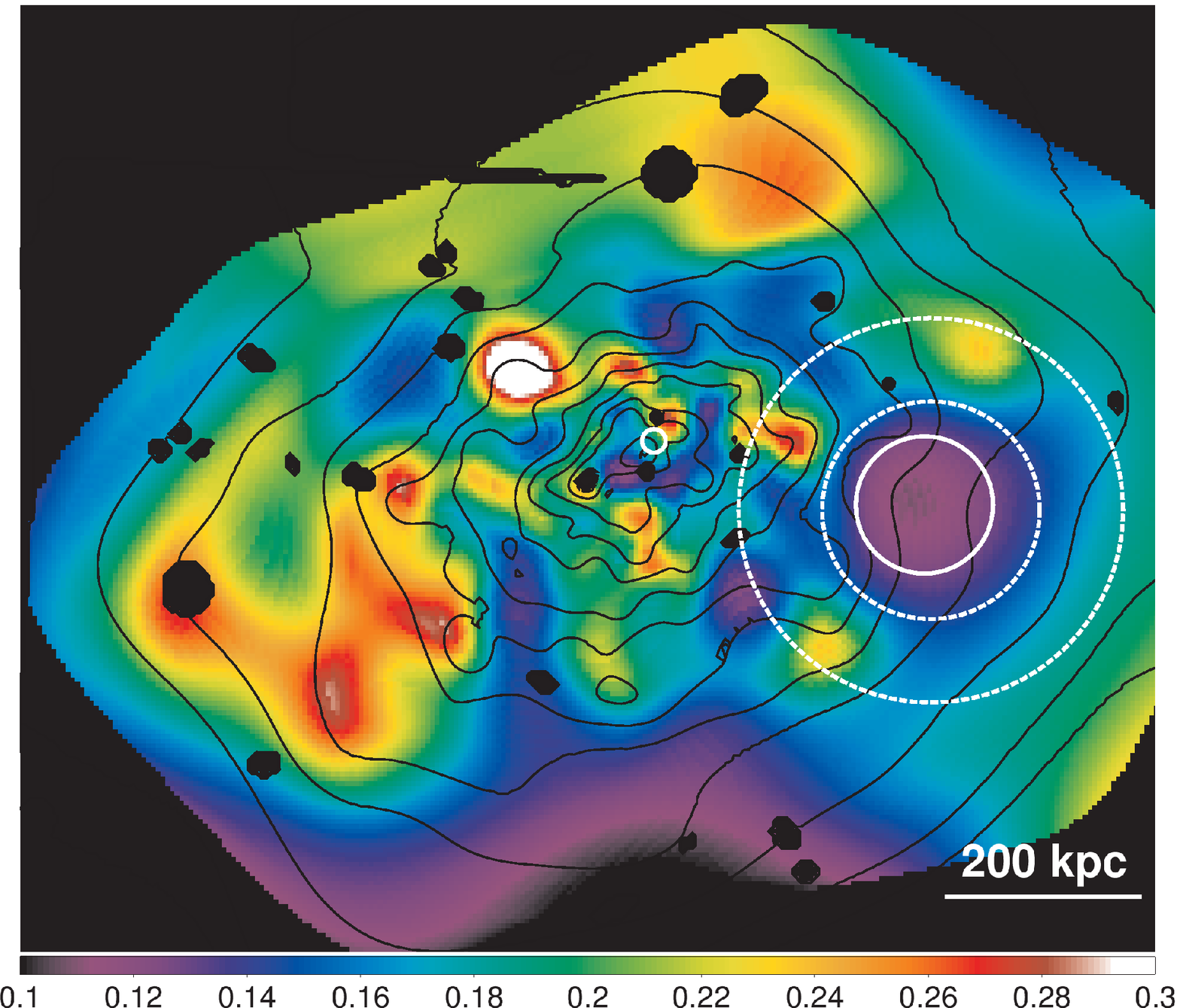}
\includegraphics[angle=0,width=.45\textwidth]{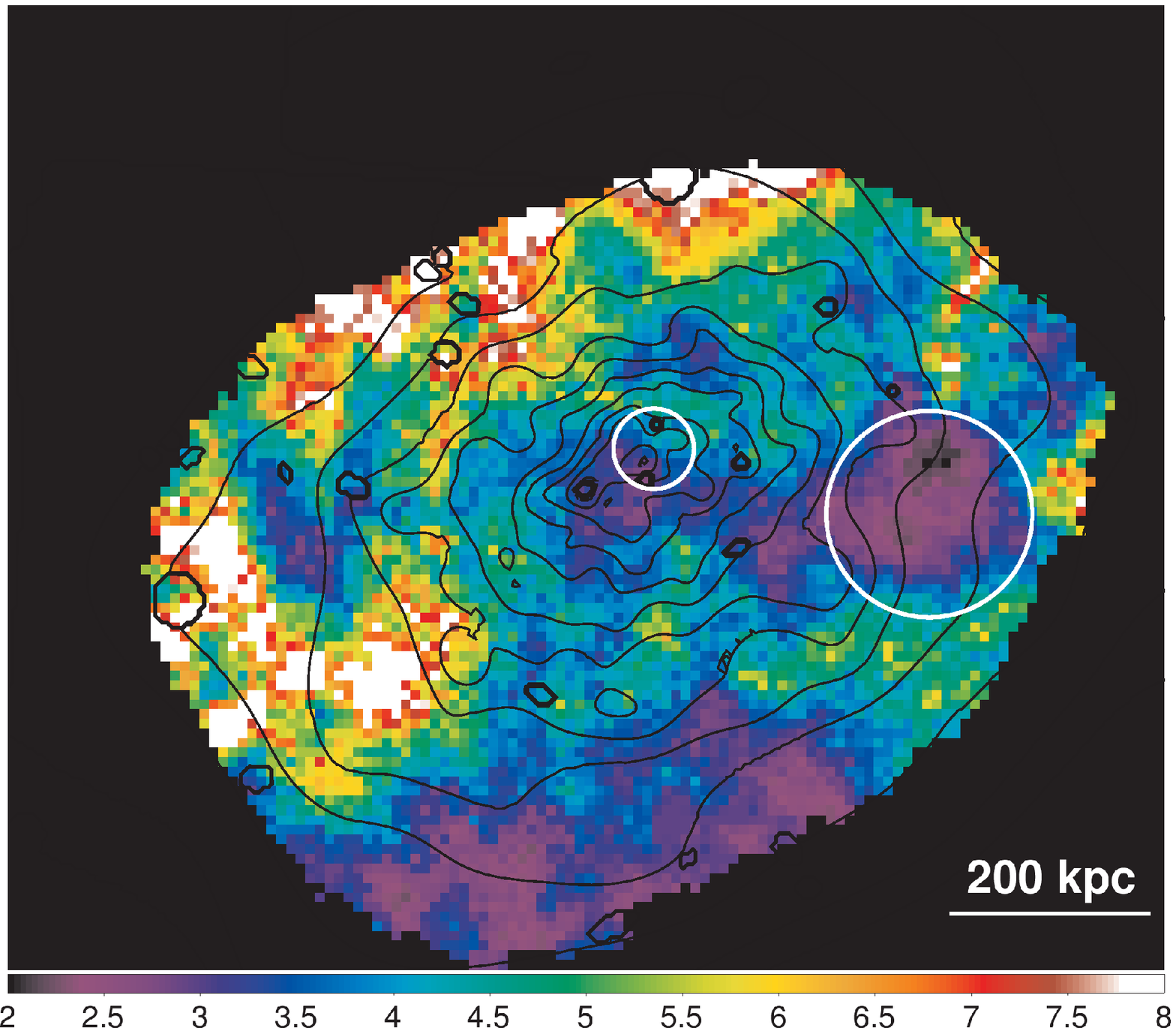}\\
\caption{Top left: An adaptively smoothed \chan\ image of A1882A. For each pixel, the $1\,\sigma$ smoothing radius is determined such that the $S/N \sim 10$ in the pixel of interest. The smoothing radius ranges from $3''$ in the brightest regions to $75''$ in the faintest regions. { The black contours show the significant positive residuals after subtraction of a smooth Beta model fitted to the cluster emission}. Top Right: SDSS RGB image of A1882A generated from the $i-, r-\, {\rm and}\, g-$band images. Bottom left: Adaptively smoothed \chan\ hardness ratio map where the $1\,\sigma$ smoothing is set such that for each pixel the relative uncertainties in the (2--5\,keV)/(0.3--2\,keV) image ratio are $20\%$. The solid circles show the $1-$sigma smoothing length ranges. { The dashed circles show the regions used to extract spectra for the soft region and its surrounds}. Bottom Right: \chan\ temperature map. The color bar shows the temperature scale in keV. The solid circles show the range in radii from which spectra are extracted. Contours from the adaptively smoothed image are overlaid onto the SDSS RGB image, hardness ratio image and temperature map. These maps do not reveal any obvious signatures of major merger activity. Both the temperature and hardness ratio maps reveal evidence for cool gas in the very central regions which may indicate the presence of a cool core.
\label{a1882a_xray_imgs}}
\end{figure*}

\begin{figure*}
\centering
\includegraphics[angle=0,width=.45\textwidth]{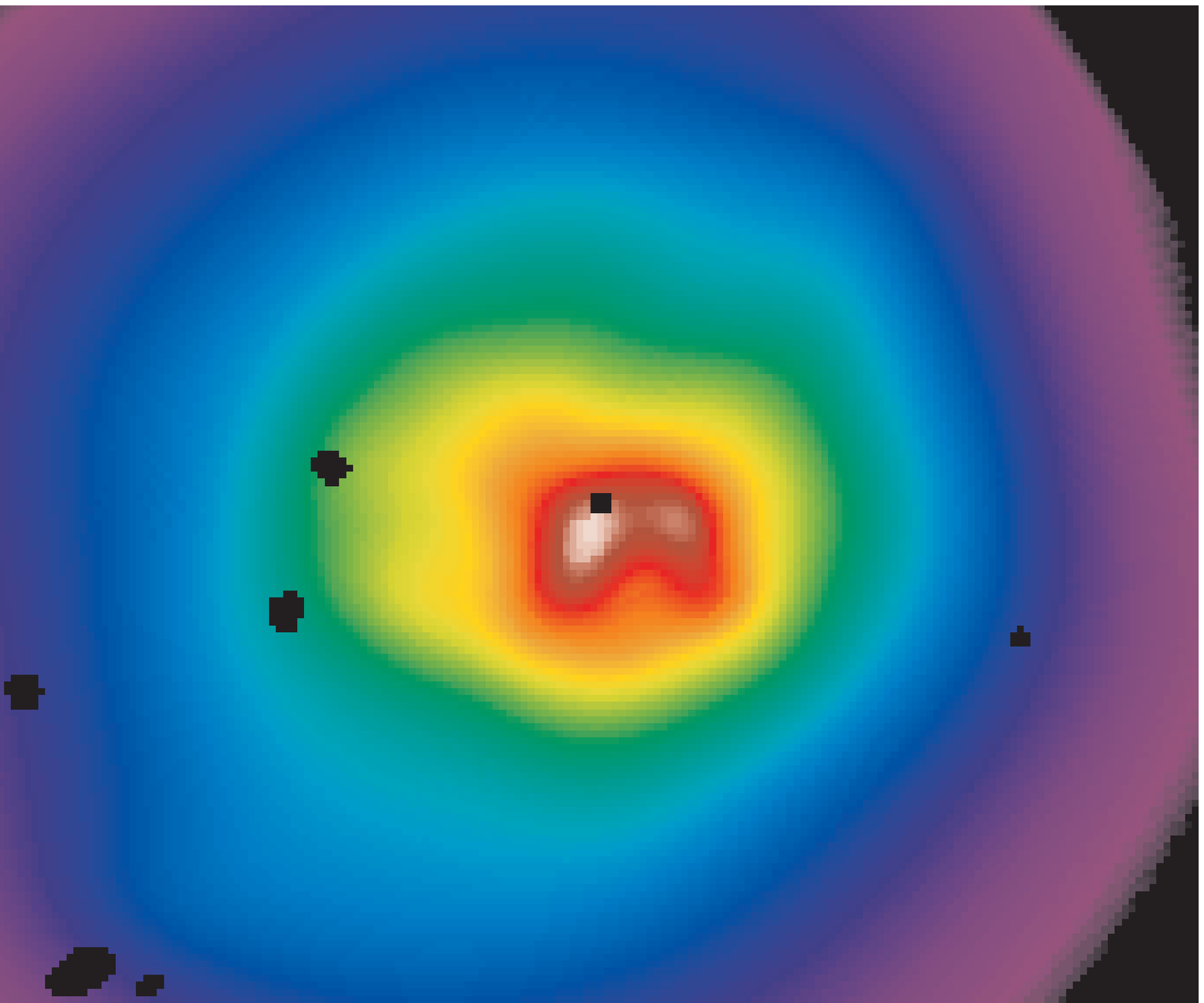}
\includegraphics[angle=0,width=.45\textwidth]{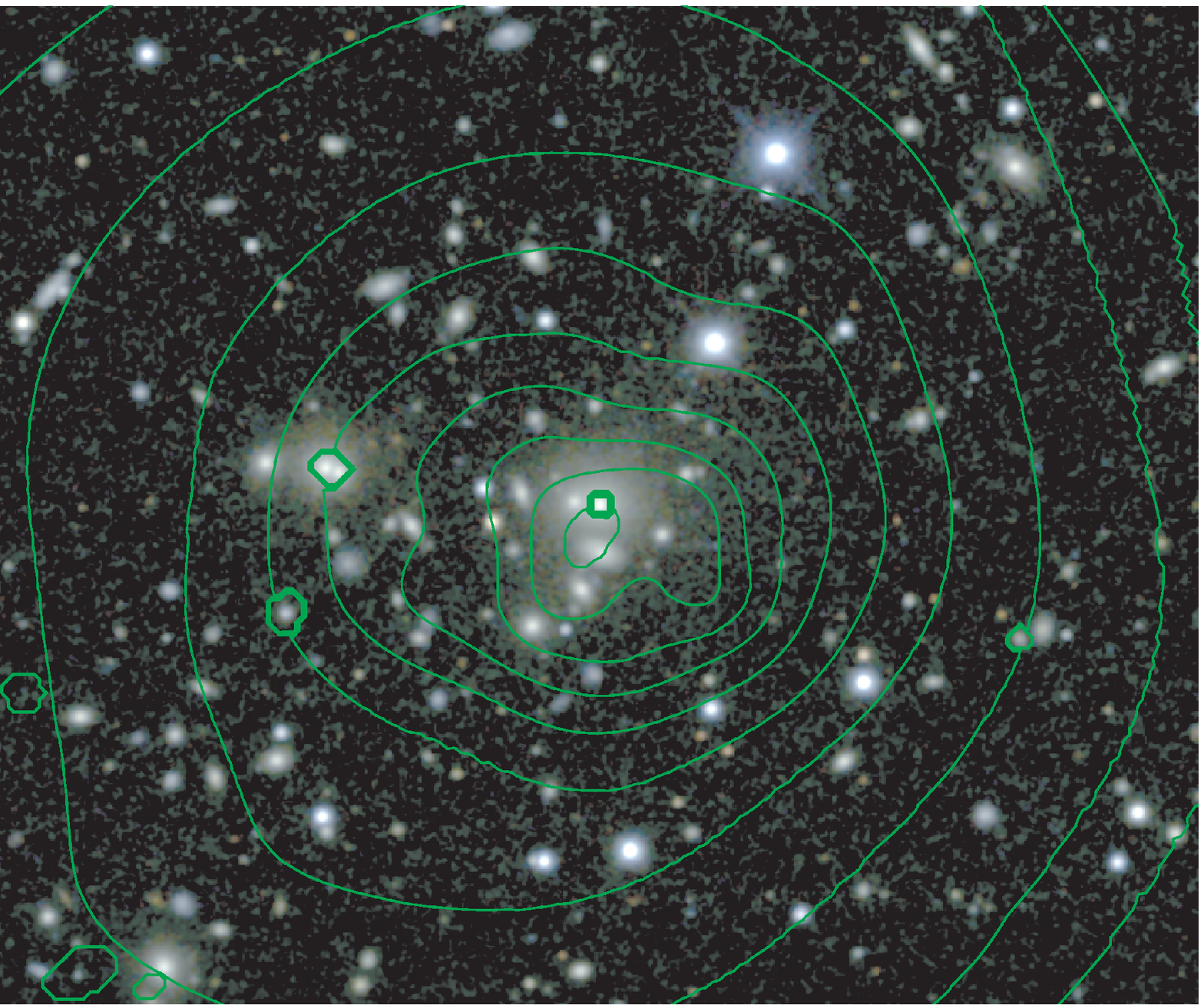}\\
\includegraphics[angle=0,width=.45\textwidth]{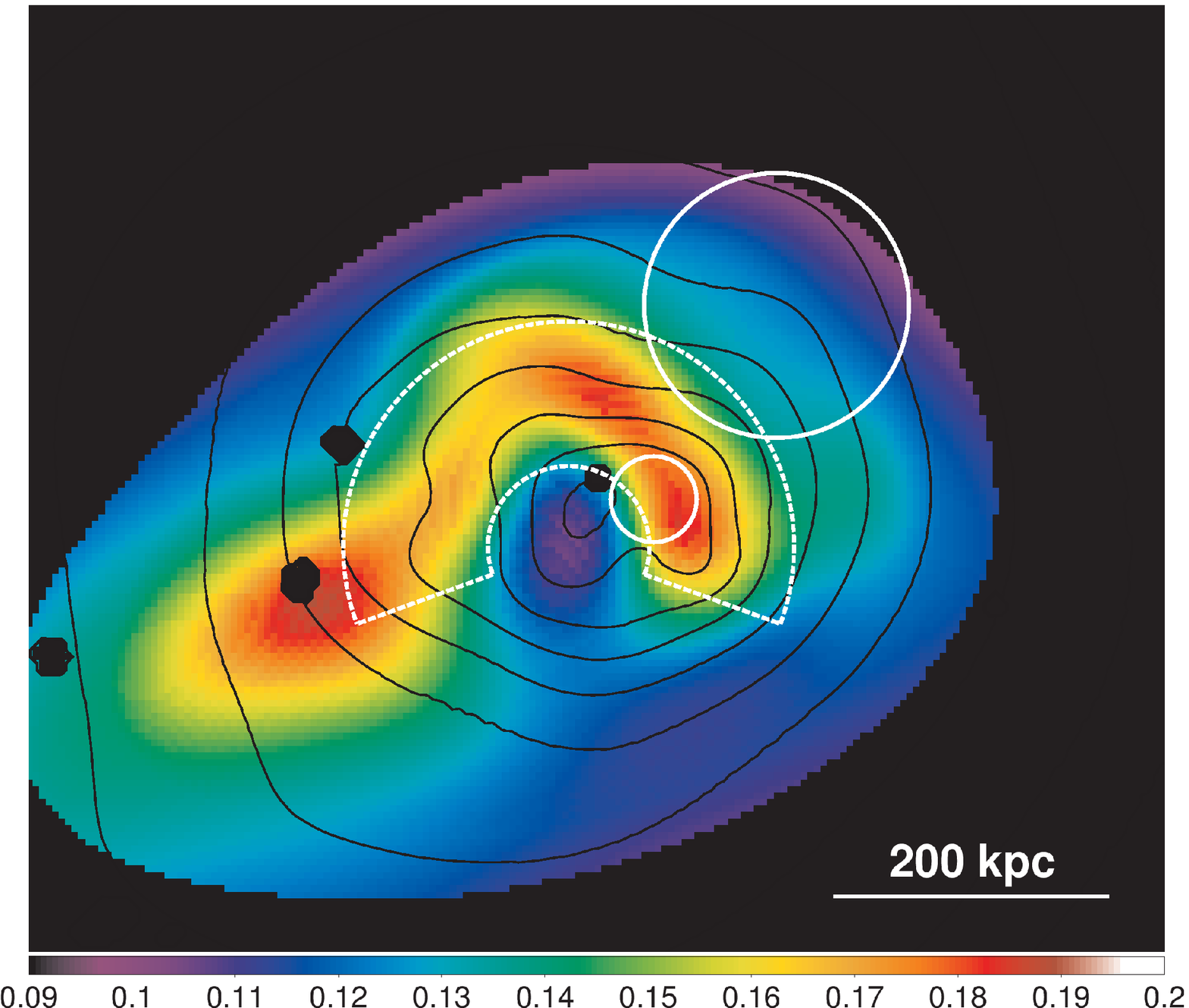}
\includegraphics[angle=0,width=.45\textwidth]{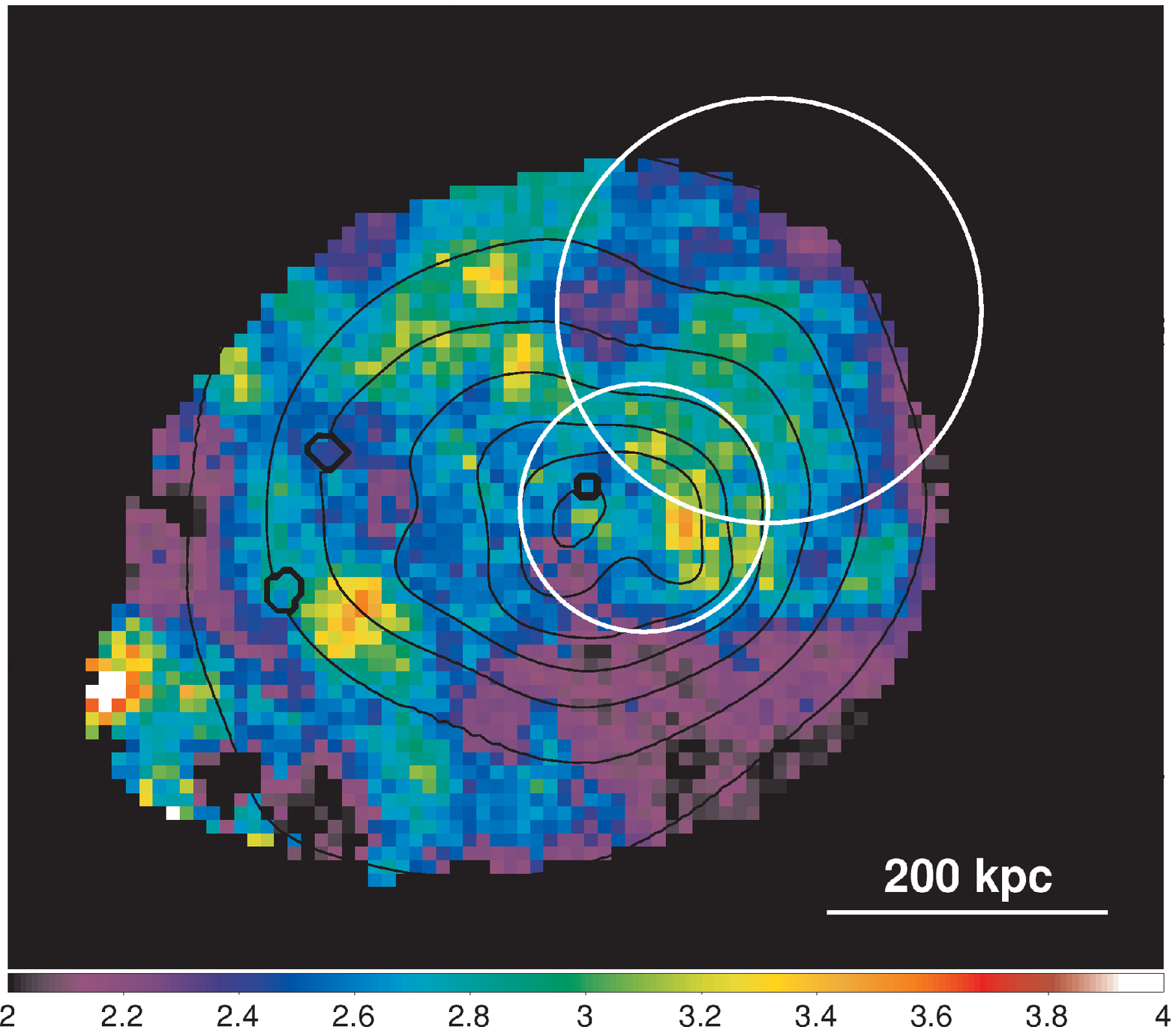}\\
\caption{Same as Figure~\ref{a1882a_xray_imgs}, but for A1882B. There does not appear to be strong evidence for cool gas in the temperature map, although this may be due to the poor spatial resolution of the map. The higher spatial resolution afforded by the hardness ratio map indicates softer emission is associated with the peak in the X-ray emission, which may indicate cool gas associated with a cool core. { The dashed white annular sector shows the region used to extract spectra which confirm that the arc-shaped region of hard emission is hotter than the mean cluster temperature.}
\label{a1882b_xray_imgs}}
\end{figure*}

In the top left panels of Figures~\ref{a1882a_xray_imgs} and \ref{a1882b_xray_imgs}, we show zoomed versions of the \chan\ images for A1882A and A1882B, respectively, which have been smoothed with an adaptive Gaussian kernel with width varying from $4'' - 79'' \simeq 10{\rm kpc} - 200\,$kpc and set such that the signal to noise in each pixel is $\sim 10$. Point sources detected with the {\sf wavdetect} software are masked during the adaptive smoothing. In the top right of Figures~\ref{a1882a_xray_imgs} and \ref{a1882b_xray_imgs} we show RGB images created from SDSS $gri$-band data with contours from the adaptively smoothed \chan\ images overlaid. The peak in X-ray emission associated with A1882A is offset from the position of the BCG by $\sim 26'' (64 {\rm kpc})$ in the direction of the ring-like distribution of galaxies to the NW. This offset may be an indication of past merger activity. Considering A1882B, the lower surface brightness and shorter exposure time compared with A1882A mean a larger smoothing scale is necessary to obtain the S/N$\sim 10$ at each pixel. At these large smoothing scales, we do not see a significant offset between the BCG and the peak in the  X-ray emission.

{ Along with the offset in the BCG position and the peak in the X-ray emission for A1882A, there appears to be a mild asymmetry to the southeast at larger radii. To highlight this feature, we use the method of \citet{neumann1997} to produce residual significance maps which highlight faint departures from a smooth Beta-model which is fitted to the data. Briefly, we use the {\it Sherpa} \citep{freeman2001} package to fit a 2D Beta-model, plus a constant background, to the surface brightness distribution of A1882A. The Beta model is defined as 
\begin{equation}
	S(r) = S_0 \left[ 1 + \left( {r} \over {r_0} \right)^2 \right]^{- \alpha} + B
\end{equation}
where $r$ is the radius which is centered at $(x_0, y_0)$, $S_0$ is the amplitude of the surface brightness and $r_0$ is the core radius. The best fitting parameters are presented in Table~\ref{beta_mod}. The model is subtracted from the data and the residual map is smoothed by a Gaussian with $\sigma=20.4'' = 50\,$kpc. This smoothed residual map is divided by an error map, which is generated assuming Poissonian statistics \citep[see][]{neumann1997}, resulting in a residual significance map. The contours from this residual significance map are overplotted onto the adaptively smoothed image shown in the top left panel of  Figure~\ref{a1882a_xray_imgs}. The contours run from $1-4\sigma$ with intervals of $1\sigma$ and show a mildly significant ($\sim 2\sigma$) positive residual southeast. There are two other enhancements of note, one just north of the X-ray peak and another to the northeast of the cluster peak. Each of the excesses coincide with features in the adaptively smoothed images. These faint residuals may also be evidence of past merger activity. We also present parameters for a Beta model fit to A1882B in Table~\ref{beta_mod}. A similar residual significance map was generated, although no notable enhancements were found. }

\begin{deluxetable*}{lcccccccc}
\tabletypesize{\scriptsize}
	\tablecolumns{7}
  \tablecaption{Parameters for a 2D Beta model fit to the \chan\ X-ray surface brightness distribution of A1882A and A1882B. \label{beta_mod}}
\tablehead{\colhead{Subcluster} &\colhead{$x_0, y_0$} & \colhead{$S_0$} & \colhead{$\alpha$} & \colhead{$r_0$} & \colhead{Ellipticity} & \colhead{Position Angle} &\colhead{$B$}\\
           & (deg., J2000)& ($10^{-7}$) & & (kpc) & & (degrees) & ($10^{-8}$)}
\startdata
A1882A & ($213.7796$, $-0.4932$) & $2.56^{+0.12}_{-0.11}$ & $1.22^{+0.15}_{-0.12}$ & $162^{+18}_{-17}$ & $0.30^{+0.02}_{-0.02}$ & $31^{+2}_{-2}$ & $7.27^{+0.14}_{-0.15}$\\
A1882B & ($213.6007$, $-0.3795$) & $1.72^{+0.18}_{-0.16}$ & $1.53^{+0.68}_{-0.39}$ & $152^{+50}_{-36}$ & $0.26^{+0.05}_{-0.06}$ & $8^{+7}_{-8}$ & $7.38^{+0.20}_{-0.26}$
\enddata
\tablecomments{The units of $S_0$ and $B$ are photons ${\rm cm}^{-2}\,{\rm s}^{-1}\, {\rm pixel}^{-2}$ where the pixel size is $3.936\arcsec \times 3.936\arcsec$. The uncertainties associated with $x_0$ and $y_0$ are $\sim 1\arcsec$ for A1882A and $\sim 2-3\arcsec$ for A1882B.}
\end{deluxetable*}

\subsection{X-ray temperatures of the substructures}\label{temps}

The mean emission-weighted temperature of the ICM is an excellent proxy for cluster mass \citep{finoguenov2001, popesso2005, vikhlinin2009}. Here, we wish to place A1882A and A1882B on the $M_{500}-T_{X, 500}$ relation of \citet{sun2009} in order to obtain an independent mass estimate for comparison with our kinematical mass measurements in Section~\ref{mass}.
We use the XMM and \chan\, data to estimate the $T_{X, 500}$, the mean X-ray temperature within the annulus $0.15r_{500}< r < r_{500}$\footnote{The central $0.15r_{500}$ region is removed to ensure that any emission associated with a cool core does not bias the temperature measurement low.}, where $r_{500}$ is the radius within which the average density is 500 times the critical density of the Universe. However, the XMM and \chan\ observations are not deep enough to trace the cluster emission to larger radii ($r > 0.5r_{500}$). Therefore, we measure the temperature within the annulus defined by $0.15r_{500} < r < r_{2500}$ to obtain $T_{X, 2500}$ where $r_{2500} \simeq 0.5r_{500}$. We then use the empirical relation  $T_{X, 500}/T_{X, 2500}=0.89$ \citep{sun2009} to extrapolate to $T_{X, 500}$.
The annuli  used for both A1882A and A1882B are shown in Figure~\ref{xray_imgs}. Point sources were removed during the extraction of the X-ray spectra. In order to facilitate a fair comparison with the kinematical masses, we use the $r_{500}$ defined in Section~\ref{mass}. 

For the XMM observations, the spectra extracted for the different cameras are fitted simultaneously with the normalizations allowed to vary and the temperatures and abundances tied. Similarly for the \chan\ observations, spectra taken with different pointings are also fitted simultaneously. The results are presented in Table~\ref{xray_temps} where it can be seen that, within the uncertainties, the temperature and abundance measurements agree well between XMM and \chan. For comparison, we also include temperature and abundance measurements with the core region included. The inclusion of the core region does not significantly affect the measured temperature.

\begin{deluxetable*}{cccc}
\tablecolumns{4}
  \tablecaption{{ Temperatures measurements for various regions of interest in A1882A and A1882B. The regions involving $r_{500}$ are shown in Figure~\ref{xray_imgs} while the regions associated with the cool spot in A1882A are shown in the lower left panel of Figure~\ref{a1882a_xray_imgs} and the region associated with the hot arc in A1882B is shown in the lower left panel of Figure~\ref{a1882b_xray_imgs}}.\label{xray_temps}} 
\tablehead {\colhead{Region} & \colhead{Temperature} & \colhead{Abundance} & \colhead{Source counts}\\
           &   \colhead{(keV)}     &   \colhead{(${\rm Z}_{\sun}$)} & \colhead{0.5--7\,keV} }
    \startdata
    A1882A (MOS1+MOS2+PN,  $r < 0.5r_{500}$) & $3.31^{+0.28}_{-0.27}$ & $0.17^{+0.12}_{-0.10}$ & 2434 \\
    A1882A (MOS1+MOS2+PN, $0.15r_{500} < r < 0.5r_{500}$) & $3.50^{+0.41}_{-0.33}$ & $0.11^{+0.14}_{-0.12}$ & 1927 \\
    A1882A (\chan\,  $r < 0.5r_{500}$) & $3.57^{+0.17}_{-0.17}$& $0.29^{+0.08}_{-0.07}$ & 10450 \\
    A1882A (\chan\,  $0.15r_{500} < r < 0.5r_{500}$) & $3.64^{+0.25}_{-0.21}$& $0.28^{+0.10}_{-0.08}$ & 8044 \\
    A1882A (\chan\,  $r < 0.15r_{500}$) & $3.50^{+0.25}_{-0.24}$& $0.32^{+0.13}_{-0.11}$ & 2847 \\
    A1882A (\chan\ cool spot) & $2.82^{+0.63}_{-0.50}$ & $0.42^{+0.37}_{-0.24}$ & 578 \\
    	A1882A (\chan\ outside cool spot) & $4.74^{+0.96}_{-0.66}$ & $0.50^{+0.40}_{-0.29}$ & 1573 \\\\
    A1882B (MOS1+MOS2+PN,  $r < 0.5r_{500}$) & $2.12^{+0.20}_{-0.20}$ & $0.25^{+0.13}_{-0.10}$ & 1349 \\
    A1882B (MOS1+MOS2+PN, $0.15r_{500} < r < 0.5r_{500}$) & $2.14^{+0.27}_{-0.25}$ & $0.31^{+0.19}_{-0.13}$ & 1030 \\
    A1882B (\chan\,  $r < 0.5r_{500}$) & $2.39^{+0.28}_{-0.28}$ & $0.21^{+0.11}_{-0.10}$ & 2178 \\
    A1882B (\chan\, $0.15r_{500} < r < 0.5r_{500}$) & $2.18^{+0.35}_{-0.25}$ & $0.13^{+0.11}_{-0.08}$ & 1603 \\
    A1882B (\chan\, $ r < 0.15r_{500}$) & $2.87^{+0.64}_{-0.36}$ & $0.53^{+0.42}_{-0.24}$ & 576 \\
    	A1882B (\chan\, hot arc) & $3.58^{+0.67}_{-0.51}$ & $0.78^{+0.47}_{-0.36}$ & 671 \\
    \enddata

\end{deluxetable*}

We also include in Table~\ref{xray_temps} results of fits to \chan\ spectra extracted from the central $0.15r_{500}$ region of A1882A and A1882B. The temperature was measured in these regions in order to search for signs of gas which is significantly cooler than the mean ICM temperature which may be associated with a cool core. Since cool cores can be destroyed during a head-on major merger, the existence of a cool core may be evidence against a recent major merger. The measured temperatures of $3.50^{+0.25}_{-0.24}\,$keV and $2.87^{+0.64}_{-0.36}\,$keV for A1882A and A1882B, respectively, are not significantly different from the mean cluster temperatures. 

\subsection{Temperature and Hardness Ratio maps for A1882A and A1882B.}
Maps of the ICM temperature in clusters often reveal evidence for past merger activity in the form of hot regions due to shocks and compression, or cool structures due to ``sloshing'' of cool core gas induced by a recent core passage \citep{ascasibar2006}. The archival \chan\ observation of A1882A has sufficient source counts to generate a temperature map with reasonable spatial resolution. To that end, we use the method described in \citet{randall2008} to generate the temperature map shown in the bottom right panel of Figure~\ref{a1882a_xray_imgs}. The method is as follows. We produce a background subtracted, combined image which is binned to $4\arcsec \times 4\arcsec$. For each pixel we generate a radius map where the radius is defined so that the circular region contains 500 0.5--7\,keV background-subtracted counts. At each pixel, we extract a source and background spectrum from a circular region defined by the radius map. The more computationally-expensive responses (ARFs and RMFs weighted by the 0.5--2\,keV flux) are produced on coarser $16\arcsec \times 16\arcsec$ grid. The spectra are fitted in XSPEC with an absorbed MEKAL model with temperature free to vary and where the hydrogen column density, redshift and abundance are fixed at $n_{\rm H}=3.22\times10^{-20} {\rm cm}^{-2}$ \citep{dickey1990},  z=0.1389 and 0.3$Z_{\sun}$, respectively. Also included in each fit is the correction for the soft X-ray background component described in Section~\ref{chandra_red}. We also present a temperature map for A1882B in the lower right panel of Figure~\ref{a1882b_xray_imgs}. However, the lower surface brightness and shorter exposure time mean that there is a high degree of correlation between the temperature measurements in each of the pixels. This is indicated by the range in extraction region size, shown as white circles in the lower right panel of Figures~\ref{a1882a_xray_imgs} and \ref{a1882b_xray_imgs}, which reveal that the extraction region radii range from $17'' - 67'' (41\,{\rm kpc} - 164\,{\rm kpc})$ and $35'' - 91'' (87\,{\rm kpc} - 222\,{\rm kpc})$ for A1882A and A1882B, respectively.

Given the poor resolution of the temperature map for A1882B, we produce the hardness ratio (HR) maps for the \chan\ observations of A1882A and A1882B which are shown in the lower left panels of Figures~\ref{a1882a_xray_imgs} and \ref{a1882b_xray_imgs}. The HR maps are produced by taking the ratios of the background subtracted, exposure corrected 2--5\,keV (hard) and 0.3--2\,keV (soft) images. The hard and soft images are smoothed with an adaptive Gaussian kernel with smoothing width set such that the relative errors on the HR at each pixel are $\sim 20\%$ \citep[a similar method is used in the adaptive binning formalism of ][]{sanders2001}. Since they require fewer source counts to obtain a significant measurement, the HR maps serve as excellent proxies for X-ray temperature but allow higher resolution maps to be produced at the expense of quantitative knowledge of the X-ray temperature \citep{henning2009}. 

Verification of the validity of the HR maps comes from comparing the HR map for A1882A with its high resolution temperature map (c.f. the temperature map for A1882B) in the lower left and right panels in Figure~\ref{a1882a_xray_imgs}, respectively. These maps reveal that the core of A1882A does in fact harbor cooler (2.5--3\,keV) gas than its immediate surrounds where the temperature increases to $\sim 4.5\,$keV. At larger radii, there are patchy regions of hot ($\gtrsim 7\,$keV) gas to the east, along with pockets of cool ($\sim2.5\,$keV) gas $\sim 120''$ to the west and south. There is a clear correlation between hard and soft regions defined in the HR map and hotter and cooler regions in the temperature map. However, the higher resolution HR maps reveal that the cool gas $\sim 120''$ west of the core is in fact not connected by a finger of cool gas to the core, as indicated in the temperature map. This is likely an artifact caused by the lower resolution of the temperature map. { We confirm that this region is cooler than its surrounds by comparing the temperature measured within a circular region encompassing the soft emission to that measured in a surrounding annular region. The spectra and responses are extracted from the \chan\ data and the regions are shown in the lower left panel of Figure~\ref{a1882a_xray_imgs}. The results are presented in Table~\ref{xray_temps} and reveal that the patch of soft emission $\sim 120''$ to the west has $kT=2.82^{+0.63}_{-0.50}$ while the surrounding gas has $kT=4.74^{+0.96}_{-0.66}$, confirming the results of the temperature and hardness ratio maps.}

While the temperature map for A1882B does not reveal any correlation with the X-ray surface brightness, the HR map clearly shows that the core region is dominated by softer emission. This softer emission may be due to either the presence of cool gas or gas with higher metallicity. The spectral fits presented in Table~\ref{xray_temps} indicate that the latter may be more likely; the temperature is not lower in the core region, but that the best-fitting metal abundance is higher than the average value, although with large uncertainties. Deeper observations are required to measure the temperatures and abundances with better precision and would help to explain the origin of the softer emission.   The HR map also reveals a striking arc-shaped region of hard X-ray emission.  The temperature maps reveal hotter ($\sim 3\,$keV) gas in the vicinity of this region, although the arc-shaped morphology is not as clear. { To confirm the temperature of this hot arc, we extract spectra and responses from the \chan\ data in the region shown in the bottom left panel of Figure~\ref{a1882b_xray_imgs}. We fit an absorbed MEKAL model to the spectra and measure a temperature of $kT=3.58^{+0.67}_{-0.51}\,$keV, consistent with the temperature map values, and confirming that this region is hotter than the mean temperature measured for A1882B.}

\section{Discussion}\label{discussion}

Based on comprehensive optical spectroscopy from the GAMA survey and archival X-ray data from both XMM and \chan\ we have detected and characterized substructure in the cluster A1882. In this section, we discuss the substructure properties and attempt to use this information to understand the ongoing merger activity. The critical question we wish to understand is whether the two main substructures have undergone, or are about to undergo, a core passage.

\subsection{The detected substructures and their masses}\label{mass}

Our analysis indicates that two substructures dominate the mass budget in the A1882 system. The first is the main structure associated with the brightest cluster member (A1882A) which has velocity dispersion $\sigma_{vpec}=500\,$\kms\ and temperature $kT=3.6\,$keV. The second, A1882B, is located $\sim 2\,$Mpc northwest of the main structure has velocity dispersion $\sigma_{vpec}=457\,$\kms, temperature $kT=2.1\,$keV and is associated with the second brightest cluster member. 
The X-ray and optical data allow the estimation of masses for A1882A and A1882B which will be used to understand the merger kinematics in Section~\ref{merger_scenario}. Optical estimates of the subcluster masses were determined using the virial estimator
\begin{equation}
M (r<r_{ap}) = M_{\rm vir}-C = {3\pi \over 2} {{\sigma_v^2 R_{PV}} \over G} - C
\end{equation}
 as defined by \citet{girardi1998} where $r_{ap}$ is the aperture radius within which we measure the mass, $\sigma_v$ is the dispersion of each cluster given in Table~\ref{kmm_fits}, $C=0.19M_{\rm vir}$ is the surface pressure correction term which allows for the cluster mass distribution external to $r_{ap}$, and 
\begin{equation}
R_{PV}={{N_{\rm ap}(N_{\rm ap}-1)} \over {\sum_{i=j+1}^{N_{\rm ap}} \sum_{j=1}^{i-1} R_{ij}^{-1}} }
\end{equation}
is the projected virial radius with $R_{ij}$ being the projected separation of the $i$th and $j$th galaxies and $N_{\rm ap}$ the number of galaxies within $r_{ap}$. For A1882A, we initially set $r_{ap}=r_{200}=\sqrt{3}\sigma_v/H(z)$ which gives the radius within which the mean density is 200 times the critical density at the cluster redshift \citep{carlberg1997}. This gives an estimate of the radius within which the cluster is virialised and, thus, the region within which it is suitable to obtain virial mass estimates. To refine our $r_{200}$ estimate, we follow the method of \citet{popesso2005} where the initial value of $M(r<r_{ap})$ is used along with an empirical estimate for the average cluster mass profile from \citet{katgert2004} to bootstrap to a new, more accurate, $r_{200}$. We iterate this process of estimating $r_{200}$ and remeasuring $M(r<r_{200})$ until the $r_{200}$ value converges. For A1882B, the initial $r_{ap}$ value is constrained to the radius of the most distant KMM-assigned member which is smaller than the $r_{200}$ estimated using its velocity dispersion. Thus, for A1882B we use the method of \citet{popesso2005} to estimate $r_{200}$ and assume an NFW profile with concentration $c=4$ to extrapolate from the $M(r<r_{ap})$ value to obtain $M(r<r_{200})$. We repeat the above procedure to determine $r_{500}$ and $M(r<r_{500})$, noting that we can directly measure $M(r<r_{500})$ for A1882B and do not rely on extrapolation. The $r_{ap}$, $r_{500}$, $r_{200}$, $M(r<r_{ap})$, $M(r<r_{500})$, and $M(r<r_{200})$ values for A1882A and A1882B are listed in Table~\ref{mass_tab2}. For A1882A, we also present equivalent masses measured using the caustics method \citep[see][for details of the method]{diaferio1999, serra2011, alpaslan2012}.

For comparison, we derive masses for A1882A and A1882B using the $M_{500}-kT$ relationship derived by \citet{sun2009}. The procedure for measuring $T_{X}$ is detailed in Section~\ref{temps}. 
The uncertainties on the $M_{500}-T_{X}$ relation, as well as those on our $T_{X}$ measurements are propagated into the final mass measurement presented in Table~\ref{mass_tab2}.
Within the uncertainties, there is good agreement between the mass measurements derived from the different methods. This consistency provides confidence that the measured masses are robust and indicate that the secondary structure, A1882B, is indeed significant being roughly half as massive as A1882A.

\begin{deluxetable*}{cccccccccc}
  \tablecolumns{10}
  \tablecaption{Estimates of mass within different radii for structures A1882A and A1882B. The values in brackets are the virial mass estimates before the surface pressure correction, $C$, is applied. \label{mass_tab2} }
 \tablehead {\colhead{Structure} & \colhead{$R_{ap}$} &\colhead{$r_{500}$} &\colhead{$r_{200}$}& \colhead{$\rm{M}_{{\rm Vir}, ap}$}&\colhead{$\rm{M}_{{\rm Vir}, 500}$} & \colhead{${\rm M}_{{\rm Caust}, 500}$} & \colhead{${\rm M}_{\rm T_{X},500}$} &\colhead{${\rm M}_{{\rm Vir}, 200}$} &\colhead{$\rm{M}_{{\rm Caust}, 200}$}\\
           & \colhead{(kpc)}  & \colhead{(kpc)} & \colhead{(kpc)} &\colhead{($\times10^{14}$\msolar)}     &   \colhead{($\times10^{14}$\msolar)}  & \colhead{($\times10^{14}$\msolar)}&\colhead{($\times10^{14}$\msolar)}     &   \colhead{($\times10^{14}$\msolar)}  & \colhead{($\times10^{14}$\msolar)}}  
    \startdata
    A1882A & 1157 & 769 & 1260 & $2.4^{+0.4}_{-0.4}(3.0^{+0.4}_{-0.4})$ & $1.5^{+0.3}_{-0.3}(1.8^{+0.3}_{-0.3})$& $1.6^{+0.5}_{-0.5}$& $1.9^{+0.2}_{-0.2}$& $2.6^{+0.4}_{-0.3}(3.2^{+0.5}_{-0.4})$& $2.6^{+0.7}_{-0.7}$\\
    A1882B &  753 & 657 & 970 &$1.0^{+0.5}_{-0.5}(1.3^{+0.6}_{-0.6})$&$0.9^{+0.5}_{-0.4}(1.1^{+0.6}_{-0.5})$ & -- & $0.8^{+0.2}_{-0.1}$ & $1.2^{+0.6}_{-0.6}$& --\\
  \enddata
\end{deluxetable*}

\subsection{Merger scenario -- post- or pre-pericentre?}\label{merger_scenario}

A key question which this study has aimed to address is the stage of the merger in A1882, i.e., are we observing a pre- or post-pericentric system? For the reasons outlined below, we assert that A1882A and A1882B have not undergone a head-on major merger in the recent past. First, the analysis presented in Section~\ref{mass} indicates that the mass ratio of A1882A and A1882B is $\sim 1:2$. Therefore, if A1882A and A1882B have recently undergone a direct head on collision it would have been quite a high-speed, violent event. The collisional nature of the ICM means that such a high-speed major merger should produce significant distortions in the X-ray morphologies of both subclusters. In addition, shocks and compression of the ICM will produce complex temperature structures observable in the temperature maps. These effects are clearly illustrated in merger simulations \citep{roettiger1996,poole2006} and observations of post-core-passage major mergers \citep{knopp1996,jones1999, markevitch2002, maurogordato2011, owers2011a}. On the contrary, the morphology and temperature structures for A1882A and A1882B (Figures~\ref{a1882a_xray_imgs} and \ref{a1882b_xray_imgs}) do not reveal {\it strong} evidence for significant recent merger activity.
Second, simulations of major head-on collisions indicate that dynamical friction significantly retards the subcluster's motion, meaning that the apocentric distances are generally much less than the virial radius ($\sim 1\,$Mpc for A1882A) \citep{tormen2004}. Therefore, the large projected separation ($\sim2\,$Mpc) of the two subclusters indicates that it is highly unlikely that A1882A and A1882B are observed after a head on collision. 
An alternative post-pericentric passage scenario which involves a less penetrative, high impact parameter merger is also unlikely. Even for high mass ratio minor mergers, the gravitational effects of the pericentric passage of a subcluster produce long-lasting  ($>\,$Gyr timescales),  easily observable ``sloshing'' cold fronts \citep{markevitch2001, ascasibar2006, markevitch2007, owers2009a, johnson2010, owers2011b, roediger2011, ma2012} in \chan\ images. More subtle low-entropy tails may also be observed in the less-massive subcluster after such an event \citep[e.g.,][]{johnson2010}. We see no evidence for such features in the X-ray data for either A1882A or A1882B. Thus, our interpretation is that A1882A and A1882B are observed in a pre-merger stage.

\begin{figure}
\centering
\includegraphics[angle=-90,width=.45\textwidth]{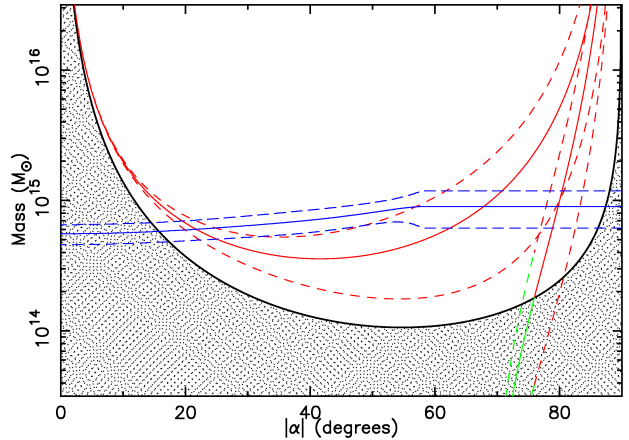}
\caption{The results of the two-body analysis showing the binding mass as a function of inclination angle, $\alpha$. The red curves show the bound solutions while the green show unbound solutions. The blue curve shows the total mass of the system enclosed with $R=R_p/cos(\alpha)$ as determined from the caustics mass profile for A1882A and the estimated $M_{200}$ for A1882B. The dashed curves show the $1\sigma$ uncertainties. Possible solutions for A1882B's orbit are found at the intersection of the blue and red/green curves. The black curve shows the region delineating bound and unbound (shaded region) solutions to the Newtonian binding criterion which indicates that A1882A and A1882B form a bound system for a large range in $\alpha$.
\label{twobody}}
\end{figure}

If A1882A and A1882B have not previously had a core crossing, it is appropriate to ask if they form a bound system and, if so, are they currently moving apart or coming together and likely to merge in the future. To that end, we perform a two-body analysis using the method of \citet{beers1982} and explained in detail in \citet{owers2009a}. The model assumes that the clusters are point sources, had zero separation at $t=0$, are either moving apart or coming together for the first time since their initial zero separation, and travel along  radial orbits. As input, the model requires the time elapsed since $t=0$, the projected separation and the LOS velocity. The projected separation, $R_p=1923\,$kpc, is the distance between the two BCGs located in the centers of A1882A and A1882B. The LOS velocity for A1882B, $V_{LOS}=-428\,$\kms\, is taken with respect to the cluster redshift and is the value determined in our KMM analysis in Section~\ref{kmm}. The time elapsed, $t=11.7\,$Gyr is the age of the Universe at z=0.1389 for our assumed $\Lambda$CDM cosmology. Given these inputs, we solve for the mass as a function of $\alpha$, the angle that the vector joining the two clusters makes with the LOS \citep[for a diagram of the assumed geometry see Figure 7. in ][]{beers1982}. In Figure~\ref{twobody} we present the solutions for the bound and unbound cases as solid red and green lines, respectively. The dashed red and green lines show the range of mass solutions due to the uncertainty in the measured  $V_{LOS}$ for A1882B, where a lower $V_{LOS}$ has bound solutions with lower masses and, conversely, bound A1882B solutions for a higher $V_{LOS}$ require a larger total mass for the system. Shown in blue is the measured total mass of the system, $M_{tot}=M^{A1882A}_{Caust}(r<R) + M^{A1882B}_{Vir,200}$. Here, we leverage the caustic technique's ability to reliably trace the mass profile of a cluster beyond the virial radius \citep{rines2012} to determine $M^{A1882A}_{Caust}(r<R)$, the mass of A1882A within radius $r<R$ where $R=R_p/cos(\alpha)$. Due to its lower mass and the difficulty in disentangling the members of the more massive A1882A in the phase-space diagram, we do not measure a caustic mass for A1882B. Instead, we simply use the virial mass reported for A1882B in Table~\ref{mass_tab2} in determining $M_{tot}$.

\begin{figure*}
\centering
\includegraphics[angle=0,width=.45\textwidth]{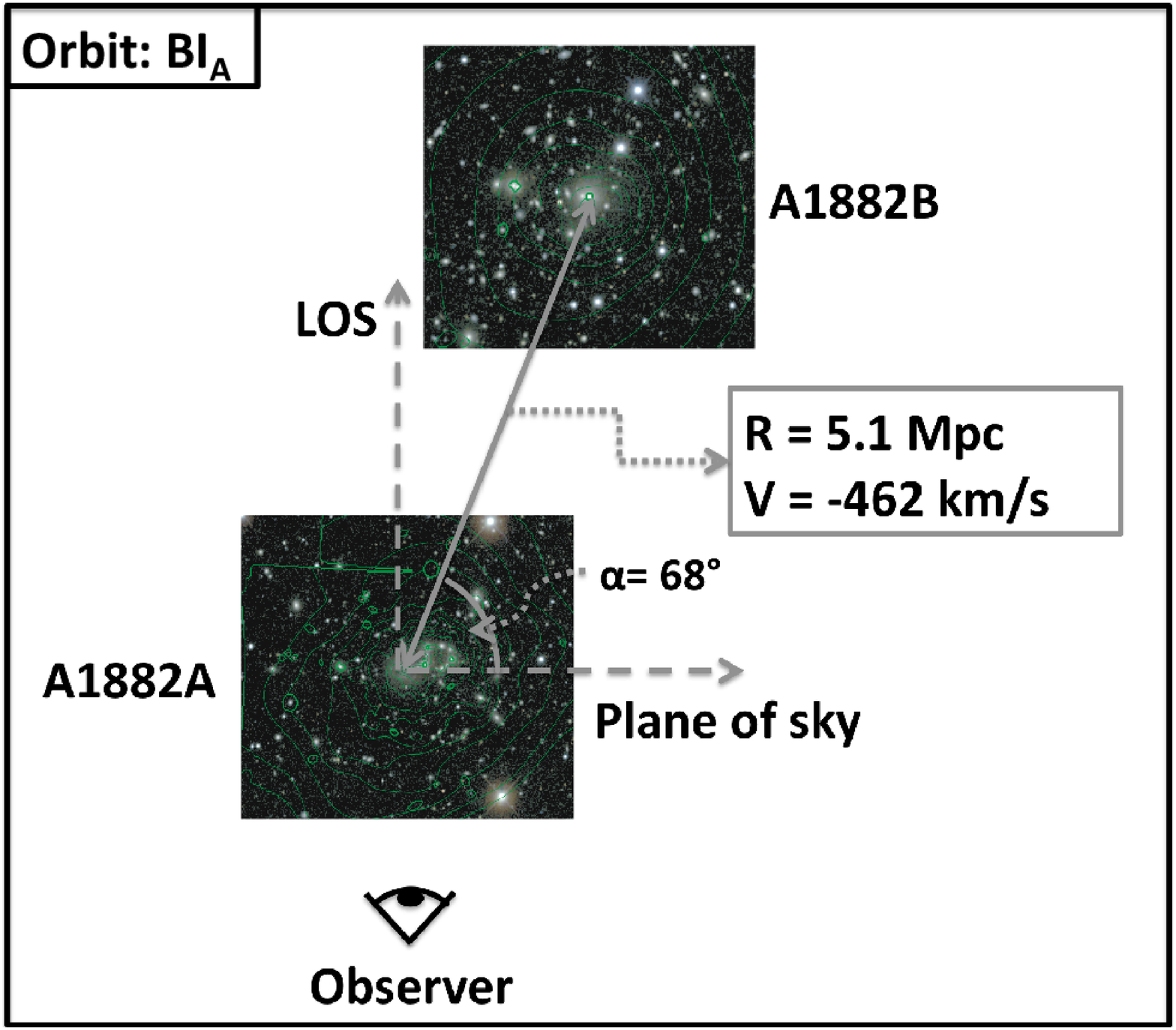}
\includegraphics[angle=0,width=.45\textwidth]{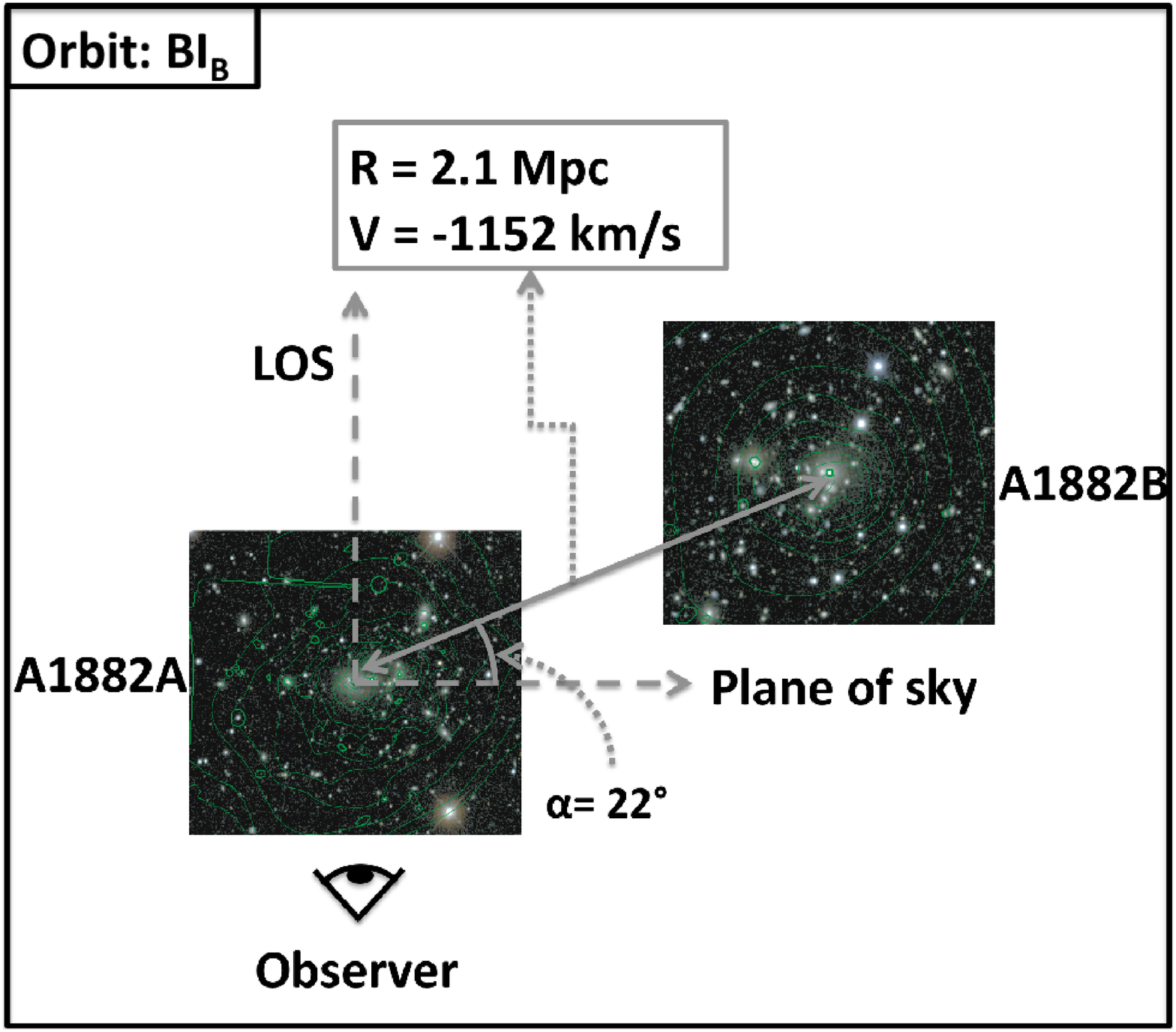}
\caption{Schematic representations for the two most probable bound and incoming orbital solutions presented in Table~\ref{twobody_probs}. The left panel shows the schematic for the $BI_A$ solution where A1882B is observed just after passing apocentre and has just begun its descent onto A1882A. The right panel shows the $BI_B$ solution where A1882B's 3D distance is very close to the observed projected separation.\label{orbits}}
\end{figure*}

The possible solutions for A1882B's orbit occurs where the blue line ($M_{tot}$) intersects the red and green lines (the bound and unbound solutions, respectively) in Figure~\ref{twobody}. At no point does the blue line intersect the green line, meaning that there are no unbound solutions for A1882B's orbit given the observed input parameters. There are three possible bound solutions for A1882B's orbit and these are listed in Table~\ref{twobody_probs}. The relative probability, $P_{rel}$, for each of the three solutions is determined as in \citet{brough2006} and relies on the assumption that the three orbital solutions are equally likely. According to this analysis, the least likely solution ($P_{rel}=5$ per cent) is that of a bound outgoing, $BO$, orbit, i.e., A1882B is observed prior to apocentric passage, is moving away from A1882A at $\sim 435\,$\kms\ along an axis within $\sim 10^{\circ}$ of our LOS and lies $\sim 10.8\,$Mpc in front of A1882A. There are two bound incoming solutions which are approximately equally probable. Schematic representations for these two solutions are presented in Figure~\ref{orbits}. The first solution (left panel Figure~\ref{orbits}), $BI_A$, places A1882B behind and traveling toward A1882A shortly after the first apocenter with $R \sim 5.1\,$Mpc, and $V \sim -462\,$\kms\, along an axis aligned to within $\sim 22^{\circ}$ of our LOS. The second solution (right panel Figure~\ref{orbits}), $BI_B$, places A1882B's orbit closer to the plane of the sky (i.e., inclined at $\sim 68^{\circ}$ to our LOS), with the distance between the two clusters, $R \sim 2.1\,$Mpc, being closer to the observed projected separation, and traveling with a higher velocity $V \sim -1152\,$\kms.

Also shown as a solid black line in Figure~\ref{twobody} is the dividing line between regions for which the Newtonian binding criterion, $V_{LOS}^2 R_P \le 2GM {\rm sin}^2 \alpha\, {\rm cos} \alpha$, holds. Solutions with masses such that they do not obey this criterion and, therefore, are unbound, are shaded. Comparing $M_{tot}$ with the Newtonian binding criterion shows that the system is bound for a large range in $\alpha$. The probability that the system is bound is $p_{bound}= 100 \times \int_{\alpha_1}^{\alpha_2} cos \alpha \, d\alpha = 73 $ per cent where $\alpha_1 = 16^{\circ}\, {\rm and\,} \alpha_2 = 87^{\circ}$ are the angles at which the observed masses (blue curve in Figure~\ref{twobody}) intersects the solid black curve.

\begin{deluxetable}{ccccc}
  \tablecolumns{3}
  \tablecaption{Solutions for A1882B's orbit as determined from the two body analysis shown inf Figure~\ref{twobody}. \label{twobody_probs}} 
\tablehead {\colhead{Orbit}   & \colhead{$\alpha$} & \colhead{$R$} & \colhead{$V$} & \colhead{$P_{rel}$} \\
& \colhead{degrees}& \colhead{Mpc}&  \colhead{\kms} & \colhead{per cent}}
\startdata
$BO$ & $-80^{+3}_{-4}$ & $10.8^{+6.8}_{-2.3}$ &$435^{+4}_{-4}$ & 4 \\ 
$BI_A$ & $68^{+13}_{-20}$ & $5.1^{+6.8}_{-2.2}$ &$-462^{+28}_{-111}$ & 50 \\ 
$BI_B$ & $22^{+11}_{-4}$ & $2.1^{+0.2}_{-0.1}$ &$-1152^{+352}_{-223}$ & 46 \\ 
\enddata

\end{deluxetable}

While we argue that A1882A and A1882B are unlikely to have had a recent encounter, there is some evidence for a dynamical disturbance in the central regions of A1882A. Namely, the offset of $64\,$kpc between the BCG and X-ray peak positions and the intriguing ring-like distribution of galaxies. { There are also subtle signatures of dynamical activity at larger radii in the form of the excess of emission to the southeast and a region of cool gas $120\arcsec$ west of the center of A1882A (Figure~\ref{a1882a_xray_imgs}). The offsets in the gas and BCG positions and the faint excess may indicate remnant bulk motions of the ICM in A1882A due to a past merger, while the cool gas may be the remnants of gas stripped during a previous merger.} However, evidence for a cool core is present in the form of gas cooler than the average ICM temperature of A1882A seen in the central part of the temperature map (Figure~\ref{a1882a_xray_imgs}). The existence of a cool core indicates that the disturbance must have been minor, since major head on mergers destroy cool cores \citep{poole2008}. Alternatively, any major interaction must have been sufficiently long ago so as to have allowed time for radiative cooling to re-establish a cool core \citep[i.e., $\gtrsim 2\,$Gyr,][]{gomez2002, poole2008}. That A1882A harbors evidence for dynamical activity which is not associated with A1882B is not surprising, particularly given the hierarchical nature of structure formation. In this sense, the A1882 system is similar to other binary clusters such as A399/A401 \citep{sakelliou2004}, A1750 \citep{belsole2004} and A1758 \citep{david2004} where the main components are observed prior to merging, while simultaneously exhibiting evidence for previous mergers.

\subsection{The cluster blue fraction}

A principal driver for this study was to understand the high blue fraction measured by \citet{morrison2003} for A1882. Given our conclusion in Section~\ref{merger_scenario} that A1882A and A1882B are observed prior to pericenter, we can rule out the effects of a major merger on the galaxies producing an enhanced blue fraction. With our spectroscopic data, we are in a position to test two further hypotheses which may explain A1882's blue fraction: (i) Is the blue fraction artificially enhanced by the contamination due to the background structure identified in Section~\ref{sec:memsel} which is nearby in redshift space but not physically associated with the cluster? (ii) Is the blue fraction anomalously high, or is it normal when compared with other clusters with like mass?

\subsubsection{Hypothesis (i): Is there contamination by the background structure?}\label{hypone}

To test the hypothesis that the background structure lying close to A1882 in redshift may be artificially enhancing the blue fraction, $f_b$ measured by \citet{morrison2003}, we compare the $f_b$ measured for spectroscopically confirmed members of A1882A and A1882B with the $f_b$ determined using a statistical background correction. If the background structure is boosting the measured cluster blue fraction, then this should be evident as a significantly higher $f_b$ measurement when using the statistical background subtraction method compared with the $f_b$ for the spectroscopically confirmed cluster members.

\begin{figure}
\centering
\includegraphics[angle=90,width=.48\textwidth]{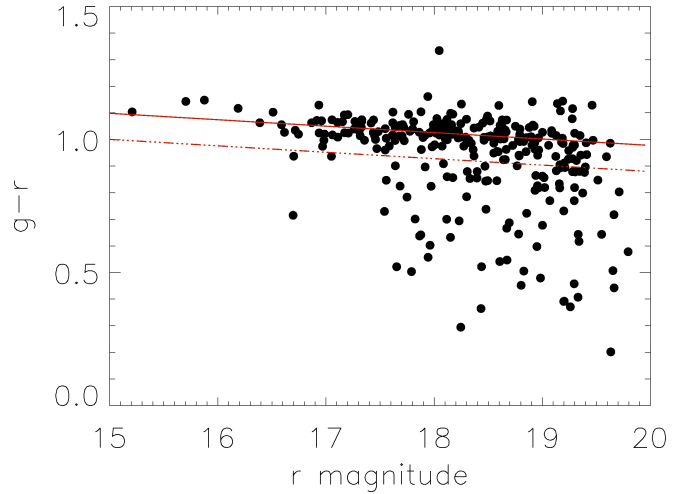}
\caption{Color-magnitude diagram for spectroscopically confirmed cluster members (filled black circles). The solid red line shows the fit to the red-sequence members and the red dot-dashed line shows the dividing line for defining blue galaxies.}.
\label{gmr}
\end{figure}

To measure $f_b$ for the spectroscopically confirmed members of A1882A and A1882B, we first need to estimate the position of the cluster red-sequence and then to use this to define blue galaxies. This is achieved by using an outlier resistant linear regression algorithm to fit a line to all spectroscopically confirmed cluster members of A1882 with $g-r > 0.9$ (this cut removes the majority of blue-cloud members) and $r \leq 19.4$. We define blue galaxies to be those that are bluer than a $2\sigma_{g-r}$ offset from the fitted cluster red-sequence where $\sigma_{g-r}$ is the standard deviation of the residuals around the fitted line. The results of the fit and the line defining blue galaxies are shown in Figure~\ref{gmr} along with the cluster member color-magnitude relation. We then determine $f_b=N_b/N_{tot}$ within the radius $r < r_{500}$ for A1882A and A1882B, where $N_b$ is the number of blue galaxies and $N_{tot}$ is the total number of galaxies. For the spectroscopically confirmed members of A1882A we find $f_b=0.28\pm0.09$ ($N_b=11, N_{tot}=40$) and similarly for A1882B we find $f_b=0.18\pm0.07$ ($N_b=7, N_{tot}=39$) where the uncertainties are calculated under the assumption of a Poissonian distribution.

To determine the $f_b$ using background subtraction, we follow a similar method to that used by \citet{urquhart2010}. We use all galaxies within $r_{500}$ and with magnitudes $15 < r <19.4$ regardless of whether they are confirmed cluster members. We measure $N_{B, clus}$ and $N_{tot, clus}$ using the definition of a blue galaxy derived above. We then define an annulus around the cluster with radius $6 - 20\,$Mpc which is to be used to determine the background corrections to $f_b$. We define 1000 regions which are randomly distributed within this annulus and with radii $3r_{500}$. We use each background region to compute a blue fraction using Equation 4 of \citet[][see also \cite{pimbblet2002}]{urquhart2010} 
\begin{equation}
f_b = {{N_{B, clus} - A N_{B, back}} \over {N_{tot, clus} - A N_{tot, back}}},
\end{equation}
where $N_{B, back}$ and $N_{tot, back}$ are the number of blue galaxies and total number of galaxies for the background regions, respectively, measured in the same manner as for the cluster regions. The background counts are scaled by $A=1/9$ which is the ratio of the cluster to background region area. For A1882A, the biweight mean of the distribution of $f_b$ values is $\langle f_b \rangle = 0.24 \pm0.03$ and for A1882B $\langle f_b \rangle = 0.23 \pm0.02$ where the uncertainties are the biweight standard deviations of the distributions and reflect the scatter in $f_b$ due to the background placement.

The two different methods for measuring $f_b$ are consistent within the uncertainties for both A1882A and A1882B.
Therefore, we conclude that the background structure has had no significant impact on the $f_b$ measurement. The blue fraction measured for the background structure (i.e., those galaxies with $1200 < v_{pec} < 3000\,$\kms\ in the top panel of Figure~\ref{vpec_rad}) is $f_b = 0.48 \pm0.1$ which is $\sim 1.5\, {\rm and}\, 2.5\,\sigma$ higher than the $f_b$ for A1882A and A1882B, respectively. Thus the blue fraction is larger in the background structure although, given its spatial distribution is different from that of the A1882 system (lower panel in Figure~\ref{vpec_rad}), it appears to have had no effect on the statistical measurements of $f_b$ for A1882A and A1882B. Based on these results, we can rule out hypothesis (i). 

\subsubsection{Hypothesis (ii): Is the blue fraction anomalously high when compared to similar clusters?}

The $f_b$ measured for A1882 in \citet{morrison2003} appears to be anomalously high when compared with other clusters of similar redshift and richness in their sample. Our analysis has shown that A1882 is comprised of two clusters with mass $\sim 10^{14}\,$\msolar. The results of \citet{urquhart2010} indicate that the cluster blue fraction shows trends with both redshift and X-ray temperature (i.e., cluster mass) in the sense that the blue fraction increases with increasing redshift and decreasing temperature. Therefore, to determine if the blue fraction in A1882 is truly anomalous, we must compare our measured values for A1882A and A1882B with blue fractions measured for clusters within a similar redshift and mass range. With that in mind, we utilize the GAMA group catalog of \citet{robotham2011} to select a sample of 46 clusters with similar redshift ($0.1 < z < 0.18$) and velocity dispersion ($300 < \sigma_v < 700\,$\kms) to A1882A and A1882B to be used as a ``benchmark'' for comparison. We use the updated, deeper GAMA redshift catalog and the caustics method (Section~\ref{sec:memsel}) to assign membership to the benchmark clusters and measured their virial masses in the same manner as was done for A1882A and A1882B in Section~\ref{mass}. We further culled the sample to contain only those benchmark clusters with virial masses in the range $6\times 10^{13} < M_{Vir, 200} < 3 \times 10^{14}$, which encompasses the range of masses allowed for A1882A and A1882B given the uncertainties on their respective mass measurements. We also cull the sample to exclude those clusters with less than 30 spectroscopically confirmed members, leaving a sample of 20 benchmark clusters. We use these benchmark clusters to produce an ensemble cluster color magnitude diagram from which we measure the blue fraction for comparison to A1882A and A1882B.

Before producing the ensemble cluster color magnitude diagram, we must ensure that we are probing the same portion of the luminosity function for the cluster galaxies across the redshift range, and that the magnitudes are $K-$corrected to the same reference frame. To that end, we use the $K-$corrections provided by \citet{loveday2012} to correct the $g-$ and $r-$band magnitudes for the A1882 and benchmark sample members to the redshift z=0.1 frame. These $K-$corrections are determined using the {\sf KCORRECT V$4\_2$} software \citep{blanton2007}. To ensure we probe the same portion of the luminosity function for all clusters, we set an absolute magnitude limit of $M_r = -19.87$. This limit is determined by the apparent magnitude limit of the spectroscopic survey ($r=19.8$) and the highest redshift cluster in the benchmark sample (z=0.1788). The position of the red sequence is determined as in Section~\ref{hypone} using the $K-$corrected $^{0.1}(g-r)$ and $^{0.1}r$ values for the A1882 cluster members. Blue galaxies were also defined as in Section~\ref{hypone} as those galaxies with $^{0.1}(g-r)$ colors bluer than  a $2\sigma_{g-r}$ offset from the fitted cluster red-sequence. 

At the brighter absolute magnitude limit, the blue fractions for the regions within $r_{500}$ of A1882A and A1882B are $f_b = 0.25\pm 0.09$ ($N_B =9, N_{tot}=36$) and $0.10 \pm 0.06$ ($N_B =3, N_{tot}=31$), respectively. For the ensemble benchmark cluster, we measure a blue fraction $f_b=0.22 \pm 0.09$ ($N_B=88, N_{tot}=406$) where the stated uncertainty is the standard deviation of the distribution of the individual benchmark cluster $f_b$ values. Within this small mass range, we see no significant difference in the blue fractions as a function of mass. Therefore, we do not attempt to split our benchmark sample based on mass in order to compare mass-matched samples to A1882A and A1882B. While the blue fraction measured for A1882A is larger than the ensemble value, it is well within the scatter of the distribution of ensemble clusters and we conclude that A1882A does not have an anomalously high fraction of blue galaxies. On the other hand, the blue fraction for A1882B is somewhat lower than the ensemble blue fraction. Given the large uncertainties and scatter in the $f_b$ measurements, however,  we conclude that there is no statistically significant difference between the ensemble and A1882B blue fractions.

\section{Summary and Conclusions}\label{summary}

We have presented a detailed analysis of the A1882 system utilizing comprehensive GAMA optical spectroscopy in combination with \chan\ and XMM X-ray data. The main findings of this analysis are:

\begin{itemize}
\item Using the combination of the highly complete, deep GAMA spectroscopy and the caustics membership selection technique we identify 283 spectroscopically confirmed cluster members. 
\item The cluster redshift is $0.1389\pm{0.0002}$ and the peculiar velocity distribution is well described by a Gaussian shape with velocity dispersion of A1882 is $525\pm23\,$\kms. This dispersion is significantly lower than previous measurements which were likely affected by the inclusion of background interlopers at slightly higher redshift.
\item The two-dimensional distribution of member galaxies reveals two major local overdensities. One is associated with the core of the main cluster, A1882A, and harbors the brightest cluster galaxy. The second, A1882B, lies $\sim 2\,$Mpc northwest of A1882A and is associated with the second brightest cluster member. Several minor  local overdensities exist in the cluster peripheries.
\item Combining the spatial and velocity information confirm A1882B as a dynamical substructure. Using the KMM algorithm to partition the data into two distinct substructures, we determine that A1882B has $v_{pec}=-428^{+187}_{-139}\,$\kms\ and $\sigma_{v}=457^{+108}_{-101}\,$\kms.
\item The \chan\ and XMM X-ray data reveal diffuse X-ray emission associated with a hot ICM coincident with both A1882A and A1882B. The mean temperature within $r_{500}$ as measured by \chan\ (XMM) is $3.57^{+0.17}_{-0.17}\,$keV ($3.31^{+0.28}_{-0.27}$) and $2.39^{+0.28}_{-0.28}\,$keV ($2.12^{+0.20}_{-0.20}$) for A1882A and A1882B, respectively.
\item The \chan\ images reveal fairly regular X-ray morphologies for both A1882A and A1882B, with no evidence for significant disturbance due to merger activity.
\item The kinematical masses agree well with the X-ray masses and indicate that A1882A ($M_{500} \sim 1.5-1.9\times 10^{14}\,$\msolar) is approximately twice as massive as A1882B ($M_{500} \sim 0.8-1.0\times 10^{14}\,$\msolar).
\item The dearth of evidence for a strong, recent head-on merger between A1882A and A1882B leads us to conclude that we are observing this system prior to merging. The Newtonian binding criterion indicates that the system has a high probability of being bound while a two-body kinematical analysis reveals that A1882B is likely bound and falling towards A1882A.
\item The fraction of blue galaxies within $r_{500}$ for both A1882A and A1882B are not anomalously high when compared with clusters of a similar mass and redshift.

\end{itemize}

Our conclusion that A1882A and A1882B are unlikely to have undergone a head-on, core passage merger in the recent past rules out a merger-related origin for the relatively high fraction of blue galaxies reported in \citet{morrison2003}. This is supported by the evidence suggesting the blue fraction is not anomalous when compared to a sample of like-mass clusters. This highlights the importance of a detailed understanding of the phase of cluster mergers, and of the masses of the involved clusters, when attempting to interpret the impact of cluster mergers on the constituent galaxies. Combining the comprehensive GAMA data presented here for a pre-merger cluster with existing data for known post-merger \citep[e.g., A1201, A3667, A2744][]{owers2009a, owers2009b, owers2011a} and relaxed clusters will allow us to assess the impact of hierarchical cluster growth on the resident galaxies.

{ We thank the referee for helping to improve the paper}. MSO acknowledges the funding support from the Australian Research Council through a Super Science Fellowship (ARC FS110200023). GAMA is a joint European-Australasian project based around a spectroscopic campaign using the Anglo-Australian Telescope. The GAMA input catalog is based on data taken from the Sloan Digital Sky Survey and the UKIRT Infrared Deep Sky Survey. Complementary imaging of the GAMA regions is being obtained by a number of independent survey programs including GALEX MIS, VST KIDS, VISTA VIKING, WISE, Herschel-ATLAS, GMRT and ASKAP providing UV to radio coverage. GAMA is funded by the STFC (UK), the ARC (Australia), the AAO, and the participating institutions. The GAMA website is www.gama-survey.org/ . This research has made use of software provided by the Chandra X-ray Center (CXC) in the application packages CIAO, ChIPS, and Sherpa and also of data obtained from the Chandra archive at the NASA Chandra X-ray Center (cxc.harvard.edu/cda/).

{\it Facilities:} \facility{CXO (ACIS)}, \facility{AAT (AAOmega)}, \facility{XMM}

\label{lastpage}

\end{document}